\documentclass[a4paper,12pt]{article}
\usepackage{epsfig}
\usepackage{epstopdf}
\usepackage[sumlimits,intlimits,namelimits]{amsmath} 
\usepackage{slashed}
\usepackage{subcaption}

\usepackage[english]{babel}
\usepackage{xcolor}
\usepackage{comment}

\usepackage{graphicx}

\usepackage{geometry}
\numberwithin{equation}{section}
\newcommand{\qcd}{\Lambda_{\rm{QCD}}}
\newcommand{\hqe}{\Lambda_{\rm{HQE}}}
\newcommand{\dv}{\Lambda_{\rm{DV}}}
\newcommand{\cdv}{\mathcal{C}_{\rm{DV}}}

\begin{document}
\begin{titlepage}
\begin{flushright}
SI-HEP-2024-15 \\ 
P3H-24-039 \\ 
Nikhef-2024-010 \\[2mm]
\today
\end{flushright}

\vspace{1.2cm}
\begin{center}
{\large\bf 
\boldmath Quark-Hadron Duality Violations and Higher-Order $1/m_b$ \\[2mm] Corrections in  Inclusive Semileptonic \boldmath $B$ \unboldmath Decays}
\end{center}

\vspace{0.5cm}
\begin{center}
{\sc Thomas Mannel} and {\sc Ilija S.\ Milutin}  \\[0.1cm]
{\sf Theoretische Physik 1,  Naturwiss. techn. Fakult\"at \\ 
Universit\"at Siegen  D-57068 Siegen, Germany} \\[0.5cm]
{\sc Rens Verkade} and {\sc K.~Keri Vos} \\[1mm]
{\sf Gravitational Waves and Fundamental Physics (GWFP), \\
Maastricht University, Duboisdomein 30, NL-6229 GT Maastricht, the Netherlands} \\ and \\
{\sf Nikhef, Science Park 105, NL-1098 XG Amsterdam, the Netherlands}
\end{center}

\vspace{0.8cm}
\begin{abstract}
\vspace{0.2cm}\noindent
The theoretical description and data for inclusive semileptonic $B$ decays have reached incredible precision. This motivated us to re-animate the discussion of possible Quark-Hadron Duality violations. There seems that there is currently no evidence of a failure of the Heavy Quark Expansion (HQE) used to compute observables for these decays. However, we might arrive 
at a point where an asymptotic behaviour of the HQE would limit a further increase of precision. 
We discuss this possibility and suggest a simple model, which can be used to study the effects of higher orders in the $1/m_b$ expansion and possible
quark-hadron duality violations. We devise observables sensitive only to such higher-order effects to test the behaviour of the HQE. Using these observables we obtain a first estimate of possible quark-hadron duality violations using the measured $q^2$ moments.

\end{abstract}

\end{titlepage}

\newpage
\pagenumbering{arabic}
\section{Introduction}
The assumption of Quark-Hadron Duality (QHD) lies at the heart of any perturbative 
QCD prediction. Starting from the relatively vague definition proposed in \cite{Poggio:1975af}
stating that a sufficiently ``smeared'' result computed at the level of quarks and gluons 
should yield the corresponding smeared quantity for the hadronic process, 
the notion of QHD has sharpened significantly \cite{Shifman:2003de,Bigi:2002fj,Bigi:2001ys,Shifman:2000jv,Blok:1997hs, Chibisov:1996wf}. 

The key to a deeper understanding of QHD is provided through the Operator Product Expansion (OPE).
It allows us to perform a systematic expansion of observables in terms of inverse powers of a 
large scale $Q$, with a numerator determined by a hadronic matrix element of the order 
$\qcd$ to the appropriate power. In this framework, QHD corresponds to a well-behaved OPE, in the best case being an analytic function of $\qcd / Q$, yielding 
a well behaved Taylor series. In turn, if QHD is violated, one expects this is not the case. 

The issue of possible QHD violations has played a significant role in the early days of heavy quark physics.  
The Heavy Quark Expansion (HQE) for inclusive processes is set up as an OPE with an expansion in powers of $\hqe / m_Q$, where 
$\hqe$ is determined by hadronic matrix elements involving heavy hadrons and is of similar size as $\qcd$, and where $m_Q$ is the mass of the heavy quark. The HQE makes use of an OPE very similar 
to the techniques used for high-energy processes, however, subleading powers of the OPE
are much more important. In the past, there have been in particular, intensive discussions 
if a precise determination of the CKM element $|V_{cb}|$ is feasible from inclusive $b \to c \ell \bar{\nu}$
decays, compared to the exclusive determination from $B \to D^* \ell \bar{\nu}$ 
where only hadronic quantities appear. 

Over the last two decades, the HQE has been refined and many theoretical issues could be 
clarified \cite{Benson:2003kp}, including the ones related to the definition of the quark masses \cite{Bigi:1996si} and the perturbative expansion. 
The HQE has been explored to order $(\hqe / m_Q)^5$ \cite{Dassinger:2006md, Mannel:2010wj, Mannel:2023yqf} and 
the perturbative expansion has been driven to N${}^3$LO for the leading term \cite{Fael:2020tow,Fael:2022frj} and NLO results 
are known for some of the subleading terms \cite{Becher:2007tk, Alberti:2012dn, Alberti:2013kxa, Mannel:2021zzr}. It is fair to say that no indications have been 
seen for a failure of the HQE, giving us confidence that a determination of $|V_{cb}|$ 
via inclusive decays is possible at a percent-level theoretical uncertainty \cite{Gambino:2016jkc,Bordone:2021oof, Bernlochner:2022ucr, Finauri:2023kte}. 

Nevertheless, there are theoretical arguments that the series in inverse powers of the heavy-quark mass 
is not analytic. Hence, one expects the presence of duality-violating contributions at some 
level. While the present phenomenology does not support large duality violations (DV) in the HQE, 
these effects can well be the limiting factor in pushing the precision of the HQE even further. 
To this end, we propose a model to constrain the size of possible QHD violations using data. This modelling of 
duality-violating contributions can be guided by the available calculations 
up to order $1/m_Q^5$. 

In order to turn all of this into a practical tool, we give a prescription of how to construct observables 
with a sensitivity to a specific order in $1/m_Q$. By studying such observables, one can constrain the size 
of higher-order terms and thereby pin down duality-violating effects. As an illustration, we discuss an 
observable constructed from $q^2$ moments, which is sensitive to terms of order $1/m_b^4$ and higher.

The paper is organized as follows. In the next section, we propose a definition of QHD violation. In Sec.~\ref{sec:T_def}, we use this definition alongside the known information about the behaviour of the HQE to guide us to the models discussed in Sec.~\ref{sec:DVModels}. Based on these models, we discuss 
the sensitivity of moments of the $b \to c \ell \bar{\nu}$ spectrum in Sec.~\ref{sec:kin_mom}. We define observables sensitive to QHDV and obtain a first extraction on the size of DV in Sec.~\ref{sec:odv}. Finally, we conclude in Sec.~\ref{sec:concl}.

\section{Definition of Duality Violations}\label{sec:def_dv}
We start by defining the OPE for the cross section for $e^+ e^- \to$ 
hadrons (discussed previously in \cite{Poggio:1975af}), which can be related at leading order in $\alpha_{\rm em}$ to the correlation function 
\begin{equation} \label{vaccor}
\Pi_{\mu \nu} (\tilde{q}) = \int \text{d}^4 x \, e^{i\tilde{q}\cdot x}
\langle 0 | T[j_\mu^{\rm em} (x)j_\nu^{\rm em} (0) ] | 0 \rangle 
= (g_{\mu \nu }\tilde{q}^2 - \tilde{q}_\mu \tilde{q}_\nu) \Pi(\tilde{q}^2) \ ,
\end{equation}
via the optical theorem 
\begin{equation}
\sigma(e^+ e^- \to {\rm hadrons}) = \frac{4 \pi \alpha_{\rm em}}{s} \;{\rm Im} \ \Pi (s)\ ,
\end{equation}
where $\tilde{q}=p_{e^+}+p_{e^-}$ and $\sqrt{s}$ is the centre-of-mass energy. The product of the two currents can be expanded for short distances $x \to 0$, corresponding
to large values of $\tilde{Q}^2 = - \tilde{q}^2$.  This yields an OPE for $\Pi (\tilde{Q}^2)$ of the form
\begin{equation} \label{ee}
\Pi (\tilde{Q}^2) = \sum_{n=0}^\infty \sum_k C_n^{(k)}
\left(\frac{1}{\tilde{Q}^2} \right)^n \langle 0 | O_n^{(k)} | 0 \rangle\ ,
\end{equation}
where the operators $O_n^{(k)}$ are of dimension $2n$ and $k$ labels the linearly independent 
operators present at dimension $2n$. This leads to the well-known condensate expansion for the 
cross section for $e^+ e^- \to$ hadrons, where the leading term corresponds to $O_0^{(1)} = 1$
as the only dimensionless operator, and $C_0^{(1)}$ is simply the partonic rate. Furthermore, 
it is well known that there are no dimension-two (gauge invariant) operators, so the expansion 
starts in this case only at order $(\qcd^2 / \tilde{Q}^2)^2$ and turns out to be very small. 

All this assumes that the OPE as a Taylor series in $1/\tilde{Q}^2$ converges to the ``real'' 
expression for the vacuum correlator in \eqref{vaccor}, which is unfortunately unknown. 
However, the presence of e.g.\ instanton 
contributions indicates that this series actually is not convergent, 
since instantons generate a dependence of the form
\begin{equation} \label{PiDV} 
    \Pi(\tilde{Q}^2) \sim \exp \left(- \omega \sqrt{\tilde{Q}^2} \right)\ ,
\end{equation}
where $\omega >0$ is a scale of order $\qcd$ related to the properties of the instanton.
This contribution cannot be expanded in $1/\tilde{Q}^2$. In the Euclidean region,
$\tilde{Q}^2 = - \tilde{q}^2 \to \infty$, 
these terms are exponentially small, but extrapolating to the Minkowskian region of positive 
$\tilde{q}^2$ generates an oscillatory behaviour, which potentially leads to a breakdown of the OPE for $\tilde{q}^2 > 0$. 
This breakdown would manifest itself as a non-convergence of the OPE, rendering it an asymptotic
series, for which the coefficients at some order $n$  in the HQE start to grow factorially. 

To quantify the meaning of this, we look at a toy example of a function $F(\lambda)$ which 
has a series representation of the form
\begin{equation} \label{Flam} 
	F(\lambda) = \sum_{n=0}^\infty a_{2n} (2n)! (\lambda^2)^n\ , 
\end{equation}
The series will converge only if the 
coefficients become factorially suppressed for large $n$, such that the $(2n)!$-term is compensated. 
However, if the series is asymptotic, 
starting at some order $n$, the $a_{2n}$ coefficients behave like constants and hence the series diverges. 

To give a meaning to \eqref{Flam}, we perform a Borel transformation:
\begin{equation}
    B[F](M) = \sum_n a_{2n} M^{2n}\ ,
\end{equation}
which now is assumed to be a Taylor series with a finite radius of convergence, although the 
original series for $F(\lambda)$ could be an asymptotic series. 
If the Taylor series of $F(\lambda)$ converges,  
the inverse of the Borel transformation is given by 
\begin{align} \label{inverse}
	 F(\lambda) =  \int\limits_0^\infty \text{d}M \, e^{-M}  B[F](\lambda M) \ ,
\end{align} 
involving an integration over $M$ along the positive real axis.  

However, if $F(\lambda)$ is an asymptotic series, the inverse cannot be calculated. To illustrate this, we assume $a_{2n} = 1$ for the asymptotic contribution to $F$ such that the asymptotic part of the Borel transform becomes 
\begin{align}
	\tilde{B}[F](M) &=  \sum_{n=0}^\infty    M^{2n} = \frac{1}{1- M^2} = \frac{1}{1+ M}\frac{1}{1- M} \ ,
\end{align} 
which exhibits a pole on the real axis at $M = \pm 1$. In this case, the inverse in \eqref{inverse} can only be determined by defining a prescription of how to deal with the singularities on the positive real axis. This prescription 
can be chosen arbitrarily and thus leads to an ambiguity in the definition of the 
inverse transform for the case of an asymptotic series. 

One convenient way to define this ambiguity is to avoid the singularity by extending 
the $M$ integration in \eqref{inverse} into the complex plane and to integrate with a 
small imaginary part $M \to M \pm i \epsilon$. The difference between the two prescriptions
defines the ambiguity, which we will identify with the duality violation. 

In this simple example, we obtain for the duality violation  
\begin{equation}\label{eq:delta}
\frac{1}{1-M+ i \epsilon} - \frac{1}{1-M- i \epsilon} = 2i \pi \delta(1-M)  \ ,
\end{equation}   
so we get (for $\lambda$ positive)
\begin{equation} \label{FlamDV}
\Delta_{\rm{DV}} F(\lambda) = 2i\pi \int\limits_0^\infty \text{d}M \, e^{-M}    
\frac{1}{1+M\lambda}\delta(1-\lambda M)  
= \frac{i \pi}{ \lambda} \exp\left(- \frac{1}{\lambda} \right) \ .
\end{equation} 
We note that this expression does not have a Taylor series in $\lambda$. Furthermore, the 
terms emerging from this ambiguity are exponentially small as $\lambda \to 0$ and thus completely 
negligible compared to the powers of $\lambda$ appearing in the series \eqref{Flam}. 

Comparing now \eqref{PiDV} with \eqref{FlamDV}, we infer
that a contribution like \eqref{PiDV} will lead to factorially growing terms in the OPE of \eqref{ee}, 
or, vice versa, if the OPE is in fact an asymptotic series, it will generate ambiguities of the form 
like in \eqref{PiDV}. 

Turning now to the HQE, we find that it has in fact a very similar structure, in particular for 
inclusive semileptonic processes. However, in this case, we use the OPE in the Minkowskian region 
since in this application we have $q^2=(p_\ell+p_\nu)^2 \sim m_Q^2$, which in the early days of the HQE raised serious 
concerns about its validity. Nevertheless, the practical application of the HQE has not 
indicated any large effects originating from such contributions. 

As discussed in more detail in the next section, the differential rate is proportional 
to a correlation function involving the hadronic $b \to c$ transition current, yielding the HQE for the 
total rate (for massless leptons)
\begin{equation} \label{b2csl}
\Gamma \propto  m_b^5 \sum_{n=0}^\infty \sum_k R_n^{(k)} (\rho) 
\left(\frac{1}{m_b} \right)^n \langle B(p) | O_n^{(k)} | B(p) \rangle\ ,
\end{equation}
where the coefficients $R_n^{(k)}$ are now functions of $\rho = m_c^2/m_b^2$, and the vacuum matrix elements 
are replaced by forward matrix elements of the decaying $B$ meson. 

Along the lines outlined above, we can now proceed to define more precisely what we call a duality 
violation in inclusive $B$ decays. It has been conjectured in \cite{Blok:1997hs} that the expansions 
\eqref{ee} and \eqref{b2csl} are in fact only asymptotic expansions, meaning that starting at some power 
the series exhibits a factorial behaviour similar to what we have discussed in the toy example.
Although we are not aware of a real proof of this assertion, the factorial growth 
of the number of independent matrix elements labelled by the index $k$ supports it. Thus, we expect on the basis of these arguments, that such contributions 
are present. 

In what follows, we discuss how to constrain 
a small duality-violating contribution in inclusive $B$ decays from the data. The problem is that the only practical tool to access inclusive semileptonic decays 
is the HQE, so we do not have any idea of the exact dependence on $m_b$ of e.g.\ the total rate. However, if the series \eqref{b2csl} is indeed asymptotic, we can use the above machinery to construct viable 
models of duality violation. Unlike the case of the vacuum correlation function \eqref{vaccor}, we apply the OPE in the 
Minkowskian region, which will result in oscillating terms instead of exponentially small ones. In fact, 
in our toy example the variable $\lambda^2$ corresponds to $1/Q^2$ such that $\lambda = 1/\sqrt{Q^2}$, where $Q^2= -q^2$. While 
in the Euclidean region $Q^2$ is positive, it will become negative in the Minkowskian case, which 
means in the toy example an analytic continuation of the form
\begin{align}
    \lambda \to i \kappa  \ , 
\end{align} 
will lead to
\begin{equation}
	\hat\Delta_{\rm{DV}} F(\lambda)     
	= \frac{\pi}{ \kappa} \exp\left(\frac{i}{\kappa} \right) \ .
 \label{eq:F_DV}
\end{equation}
Finally, the decay rate is obtained by taking the imaginary part, so we end up (schematically) with 
\begin{equation}
\Gamma_{\rm{DV}} \sim \sin \left(\frac{m_b}{\dv} \right) \ ,
\end{equation}
where $\dv$ is a scale related to DV, which one would expect to be of order $\qcd$.

\section{Modelling Duality Violations using the HQE} \label{sec:T_def}
The HQE  in inclusive semileptonic decays $B(p)\to X_c(p_X)\ell (p_\ell)\bar{\nu}(p_\nu)$ is set up using the 
optical theorem. In order to obtain differential rates, the starting point is the correlation 
function of two hadronic currents
\begin{equation}
T_{\mu \nu} (q) = \int \text{d}^4 x \, e^{iq\cdot x} \langle B (p) | 
T\{ \bar{b}(x) \Gamma_\mu   c(x) \, \bar{c}(0) \overline{\Gamma}_\nu   b(0) \} | B(p) \rangle   \ ,
\end{equation}
with $\Gamma_\mu= \frac{1}{2}\gamma_\mu (1-\gamma_5)$ and $q^\mu=(p_\ell+p_\nu)^\mu$. 
Contracting this with the leptonic tensor yields the forward scattering amplitude, 
the imaginary part of which is the inclusive rate. 

In order to set up the HQE, it is convenient to re-define the heavy quark field according to  
\begin{equation}
b(x) = \exp (-i m_b\,  v\cdot x) b_v(x)\ , \qquad v=p/M_B\ ,
\end{equation} 
where $v$ is the four velocity of the decaying heavy meson. This re-definition  
corresponds to a splitting of the heavy-quark momentum $p_b = m_b v + k$, where $k$ is a small 
residual momentum related to the covariant derivative acting on $b_v(x)$. This yields     
\begin{equation} \label{Tprod1} 
T_{\mu \nu} (Q) = \int \text{d}^4 x \, e^{-iQ\cdot x} \langle B (p) | 
T\{ \bar{b}_v(x) \Gamma_\mu c(x) \, \bar{c}(0) \overline{\Gamma}_\nu b_v(0) \} | B(p) \rangle \ ,  
\end{equation}
where $Q =  m_b v - q$ .
The hadronic correlation function can be decomposed into five scalar functions $T_i$ 
\begin{align}
T_{\mu \nu} (Q) =&\ T_1 \left(g_{\mu \nu} + \frac{Q_\mu v_\nu + Q_\nu v_\mu - i \epsilon_{\mu \nu \alpha \beta} Q^\alpha v^\beta}{vQ} \right)\nonumber\\ 
& - T_2 g_{\mu \nu} + T_3 v_\mu v_\nu 
+ T_4 \frac{(Q_\mu v_\nu + Q_\nu v_\mu )}{vQ} + T_5 \frac{Q_\mu Q_\nu}{(vQ)^2}\ ,\label{eq:T_decomp}
\end{align}  
where the scalar functions depend on 
\begin{equation}
T_i \equiv T_i (vQ, Q^2) \ .
\end{equation} 

The tree level expression of the HQE in \eqref{Tprod1} can be obtained by inserting the external field 
propagator for the charm quark \cite{Dassinger:2006md}.  Expanding the external field propagator gives\footnote{In the following, we neglect the mass of the charm quark in order to simplify the discussion.}
\begin{align}
    \frac{1}{\slashed Q+i\slashed D}&=\sum\limits_{k=0}^\infty \Big(\frac{1}{Q^2}\Big)^{k+1}\slashed Q\big[-(i\slashed D)\slashed Q\big]^k\ .
\end{align}
Inserted this into \eqref{Tprod1} and taking the forward matrix element with a $B$ meson 
state moving with velocity $v$ yields 
\begin{equation} \label{eq:Tmunu}
T_{\mu \nu} (Q) =  \sum_{k=0}^\infty \left(\frac{1}{Q^2} \right)^{k+1} 
\langle B (v) | \bar{b}_v \Gamma_\mu \slashed{Q} [ -(i \slashed{D}) \slashed{Q}]^k \overline{\Gamma}_\nu b_v(0) | B(v) \rangle \ .
\end{equation} 

We can now project out the scalar components $T_i (vQ,Q^2)$, by contracting the indices with 
appropriately chosen tensors. The forward matrix elements then become functions of $vQ$ and $Q^2$ to some power depending on the number of covariant derivatives. Schematically, we have for the first three terms in the sum
\begin{align}
 \langle B(v) | \bar{b}_v \Gamma \slashed{Q} \overline{\Gamma} b_v | B(v) \rangle &= a_0^{(i,0)} (vQ)\  \nonumber \\
 \langle B(v) | \bar{b}_v (-1) \Gamma \slashed{Q} (i \slashed{D}) \slashed{Q} \overline{\Gamma}  b_v 
| B(v) \rangle &= \hqe \left( a_0^{(i,1)} (vQ)^2 + a_1^{(i,1)} Q^2 \right)\  \nonumber \\ 
 \langle B(v) | \bar{b}_v \Gamma \slashed{Q} (i \slashed{D}) \slashed{Q} (i \slashed{D}) \slashed{Q}   \overline{\Gamma} b_v 
| B(v) \rangle &= \hqe^2 \left( a_0^{(i,2)} (vQ)^3 + a_1^{(i,2)} (vQ) Q^2 \right) \ ,
\end{align}
where the $i$ on the coefficients indicates the scalar component of the gamma matrices as in \eqref{eq:T_decomp}. 

In general, we thus have
\begin{align}
 \langle B(v) | \bar{b}_v \Gamma \slashed{Q} [ -(i \slashed{D}) \slashed{Q}]^k \overline{\Gamma} b_v 
| B(v) \rangle & = \hqe^k   \sum_{j=0}^{j_{\rm max}} a_j^{(i,k)} (vQ)^{k-2j+1} (Q^2)^{j} \ ,
\label{eq:traceform}
\end{align}
where $j_{\rm max} = (k+1)/2$ and $j_{\rm max}=k/2$ for odd and even $k$, respectively. 
To investigate this structure, it is useful to introduce the variables 
\begin{equation}
r^2 \equiv \frac{Q^2}{\hqe^2} \quad \mbox{and} \quad t \equiv\frac{vQ}{\hqe}  \ .
\end{equation}    
Inserting the expressions in \eqref{eq:Tmunu} and collecting terms with equal powers of $r$ then gives
\begin{equation} \label{TAnsatz}
	T_i (t,r^2) = \frac{1}{\hqe}  \sum_{l=0}^\infty \left(\frac{1}{r^2} \right)^{l+1} 
	P_l^{(i)} (t)\ ,
\end{equation}  
where $P_l^{(i)} (t)$ is a polynomial of order $l+1$ in $t$:
\begin{equation} 
	P_l^{(i)}(t) = \sum_{n=0}^{l+1} t^{l+1-n} a_n^{(i,n+l)}\ . 
\end{equation}
Using the trace formula from \cite{Mannel:2023yqf}, we can calculate these coefficients at tree-level in terms of the HQE elements up to $1/m_b^5$. The HQE elements are defined in Appendix~\ref{app:inputs}. In Table \ref{tab:a_coeff}, we present the leading HQE contribution to $a_n^{(i,n+l)}$ in terms of $\mu_G^2, \mu_\pi^2$ and $\tilde{\rho}_D^3, \rho_{LS}^3$ coefficients of dimension 5 and dimension 6 respectively. The coefficients $a_0^{(i,0)}$ correspond to the partonic result and therefore do not receive any contributions when including higher order corrections in the $1/m_b$ expansion. Moreover, the leading contribution to the coefficients $a_n^{(i,1)}$ are of order $\hqe/m_b$ instead of order 1 like the other coefficients, since in the HQE the $1/m_b$ contribution vanish due to heavy quark symmetries. 
\renewcommand{\arraystretch}{1.2}
\begin{table}[t!]
\begin{center}
    \begin{tabular}{|c||c|c|c|c|c|}
         \hline
          & \boldmath$T_1$ \unboldmath &\boldmath $T_2$\unboldmath &\boldmath $T_3$\unboldmath & \boldmath$T_4$\unboldmath & \boldmath$T_5$ \unboldmath\\
         \hline
         \hline
         
         \boldmath $a_0^{(i,0)}$ \unboldmath & $-\frac{1}{2}$ &0 & 0&1 & 0\\
         
         \hline
         
         \boldmath $a_0^{(i,1)}$ \unboldmath & $-\frac{5}{6}\left(\frac{\mu_G^2-\mu_\pi^2}{m_b\hqe}\right) $ & 0 & 0 &$\frac{5}{3}\left(\frac{\mu_G^2-\mu_\pi^2}{m_b\hqe}\right) $& $-\frac{2}{3}\left(\frac{\mu_G^2-\mu_\pi^2}{m_b\hqe}\right) $\\
         
         \boldmath $a_1^{(i,1)}$ \unboldmath &0 &$-\frac{5}{12}\left(\frac{\mu_G^2-\mu_\pi^2}{m_b\hqe}\right) $ & $-\frac{5}{6}\left(\frac{\mu_G^2-\mu_\pi^2}{m_b\hqe}\right) $& 0& 0\\
         
         \hline
         
         \boldmath $a_0^{(i,2)}$ \unboldmath & $-\frac{2}{3}\frac{\mu_\pi^2}{\hqe^2} $ &0 &0 &$\frac{4}{3}\frac{\mu_\pi^2}{\hqe^2} $ &$\frac{4}{3}\left(\frac{\tilde{\rho}_D^3+\rho_{LS}^3}{m_b\hqe^2}\right) $ \\
         
         \boldmath $a_1^{(i,2)}$ \unboldmath &$-\frac{1}{6}\left(\frac{\mu_G^2+\mu_\pi^2}{\hqe^2}\right) $ & $-\frac{1}{3}\frac{\mu_\pi^2}{\hqe^2}$&$-\frac{2}{3}\frac{\mu_\pi^2}{\hqe^2} $ & $\frac{1}{3}\frac{\mu_\pi^2}{\hqe^2} $& 0 \\
         
         \hline
         
         \boldmath $a_0^{(i,3)}$ \unboldmath & $-\frac{4}{3}\frac{\tilde{\rho}_D^3}{\hqe^3} $ &0 &0 &$\frac{8}{3}\frac{\tilde{\rho}_D^3}{\hqe^3} $ & 0\\ 
         
         \boldmath $a_1^{(i,3)}$ \unboldmath &$\frac{2}{3}\frac{\tilde{\rho}_D^3}{\hqe^3} $ &$-\frac{2}{3}\frac{\tilde{\rho}_D^3}{\hqe^3} $ &$-\frac{4}{3}\frac{\tilde{\rho}_D^3}{\hqe^3} $ & $-\frac{2}{3}\left(\frac{2\tilde{\rho}_D^3-\rho_{LS}^3}{\hqe^3}\right) $& $-\frac{2}{3}\frac{\rho_{LS}^3}{\hqe^3} $\\
         
         \boldmath $a_2^{(i,3)}$ \unboldmath &0 &$\frac{1}{6}\left(\frac{3\tilde{\rho}_D^2-\rho_{LS}^3}{\hqe^3}\right) $ & $\frac{1}{3}\left(\frac{3\tilde{\rho}_D^2-\rho_{LS}^3}{\hqe^3}\right) $&0 & 0\\
         
         \hline
    \end{tabular}
\end{center}
  \caption{ The coefficients $a_n^{(i,n+l)}$, defined in \eqref{eq:traceform}, in terms of the non-perturbative HQE parameters (see Appendix \ref{app:inputs}).  We present here only the leading contributions in terms of dimension 5 or 6 HQE parameters, dropping corrections of higher dimensions of order $\mathcal{O}(\hqe/m_b)$.}
 \label{tab:a_coeff}
\end{table}
\renewcommand{\arraystretch}{1}

We can now construct a model for duality violation based on the discussion of 
Sec.~\ref{sec:def_dv}. We make the ansatz for the polynomials $P_l^{(i)} (t)$ in \eqref{TAnsatz} to be
of the form 
\begin{align} \label{pdef}
    P_l^{(i)}(t)&= (2l)! p_l^{(i)}(t)=(2l)!\sum\limits_{n=0}^{l+1}t^{l+1-n}b_n^{(i,n+l)}\ ,
\end{align}
where $b_n^{(i,n+l)} = a_n^{(i,n+l)}/(2l)!$  such that the redefined polynomial $p_l^{(i)}(t)$ remains of the same magnitude for growing $l$. 
This yields
\begin{align}
    T_i(t,r^2)&=\frac{1}{\hqe}\frac{1}{r^2}\sum\limits_{l=0}^{\infty}\left(\frac{1}{r^2}\right)^l (2l)!p_l^{(i)}(t)\ . \label{eq:Fakt}
\end{align}
In order to proceed further, we need an assumption about the coefficients of $p_l^{(i)}(t)$, 
which eventually defines the model. There are various ways to discuss the $t$ dependence of a viable 
model. Here we use the explicit calculation of the $a_n^{(i,j)}$ coefficients in the HQE up to $1/m_b^5$ to guide the modelling of $p_l$.
In Table~\ref{tab:a_coeff}, we already listed the exact expressions up to $1/m_b^3$. However, for a quantitative analysis, we need numerical values for the HQE parameters. Since the 
HQE parameters have been fitted to data only up to $1/m_b^3$ \cite{Bordone:2021oof,Finauri:2023kte} and partially to $1/m_b^4$ \cite{Bernlochner:2022ucr}, we employ the ``lowest-lying state saturation ansatz'' (LLSA) \cite{Heinonen:2014dxa} to obtain numerical estimates for the higher-order 
HQE parameters. We present the numerical values of the coefficients $b_n^{(i,n+l)}$ 
to $\mathcal{O}(1/m_b^5)$ in 
Table \ref{tab:b_coeff}. All input values are given in Appendix \ref{app:inputs} and we use $\hqe= 0.5$ GeV.

\begin{table}[t!]
\centering
\begin{subtable}{.5\textwidth}
\centering
\begin{tabular}{|c||r|r|r|r|}
\hline
\multicolumn{5}{|c|}{\boldmath $T_i$ \unboldmath} \\
\hline
         \boldmath $l=0$ \unboldmath & $b_0^{(i,0)}$ & $b_1^{i,1}$ & - & - \\
         
         \boldmath $l=1$ \unboldmath & $b_0^{(i,1)}$ & $b_1^{i,2}$ & $b_2^{i,3}$ & - \\
         
         \boldmath $l=2$ \unboldmath & $b_0^{(i,2)}$ & $b_1^{i,3}$ & $b_2^{i,4}$& $b_3^{i,5}$ \\
         
         \boldmath $l=3$ \unboldmath & $b_0^{(i,3)}$ & $b_1^{i,4}$ & $b_2^{i,5}$& $\mathcal{O}(1/m_b^6)$ \\
         
         \boldmath $l=4$ \unboldmath & $b_0^{(i,4)}$ & $b_1^{i,5}$ & $\mathcal{O}(1/m_b^6)$& $\mathcal{O}(1/m_b^6)$ \\
         
         \boldmath $l=5$ \unboldmath & $b_0^{(i,5)}$ & $\mathcal{O}(1/m_b^6)$ & $\mathcal{O}(1/m_b^6)$&$\mathcal{O}(1/m_b^6)$ \\
         
         \hline
\end{tabular}

\end{subtable}\\ \vspace{0.5cm}
\begin{subtable}{.5\textwidth}
\centering
\begin{tabular}{|c||r|r|r|r|}
\hline
\multicolumn{5}{|c|}{\boldmath $T_1$ \unboldmath} \\
\hline
         \boldmath $l=0$ \unboldmath & -0.5 & 0 & - & - \\
         
         \boldmath $l=1$ \unboldmath & 0.032 & -0.265 & 0 & - \\
         
         \boldmath $l=2$ \unboldmath & -0.052 & 0.050 & 0.002 & 0 \\
         
         \boldmath $l=3$ \unboldmath & -0.003 & 0.001 & -0.0005&$\mathcal{O}$ \\
         
         \boldmath $l=4$ \unboldmath & -0.0002 & 0.0004 &$\mathcal{O}$& $\mathcal{O}$ \\
         
         \boldmath $l=5$ \unboldmath & -0.000007 & $\mathcal{O}$ &$\mathcal{O}$&$\mathcal{O}$ \\
         
         \hline
\end{tabular}

\end{subtable}\\ \vspace{0.5cm}
\begin{subtable}{.5\textwidth}
\centering
\begin{tabular}{|c||r|r|r|r|}
\hline
\multicolumn{5}{|c|}{\boldmath $T_2$ \unboldmath} \\
\hline
         \boldmath $l=0$ \unboldmath & 0 & 0.032 & - & - \\
         
         \boldmath $l=1$ \unboldmath & 0 & -0.310 & 0.570 & - \\
         
         \boldmath $l=2$ \unboldmath & 0 & -0.043 & 0.049 & 0.031 \\
         
         \boldmath $l=3$ \unboldmath & 0 & -0.005 & 0.017&$\mathcal{O}$ \\
         
         \boldmath $l=4$ \unboldmath & 0 & -0.0003 &$\mathcal{O}$& $\mathcal{O}$ \\
         
         \boldmath $l=5$ \unboldmath & 0 & $\mathcal{O}$ &$\mathcal{O}$&$\mathcal{O}$ \\
         
         \hline
\end{tabular}

\end{subtable}
\begin{subtable}{.5\textwidth}
\centering
\begin{tabular}{|c||r|r|r|r|}
\hline
\multicolumn{5}{|c|}{\boldmath $T_3$ \unboldmath} \\
\hline
         \boldmath $l=0$ \unboldmath & 0 & 0.064 & - & - \\
         
         \boldmath $l=1$ \unboldmath & 0 & -0.620 & 1.119 & - \\
         
         \boldmath $l=2$ \unboldmath & 0 & -0.086 & 0.154 & 0.015 \\
         
         \boldmath $l=3$ \unboldmath & 0 & -0.010& 0.036&$\mathcal{O}$ \\
         
         \boldmath $l=4$ \unboldmath & 0 & -0.0006 &$\mathcal{O}$& $\mathcal{O}$ \\
         
         \boldmath $l=5$ \unboldmath & 0 & $\mathcal{O}$ &$\mathcal{O}$&$\mathcal{O}$ \\
         
         \hline
\end{tabular}

\end{subtable}\\ \vspace{0.5cm}
\begin{subtable}{.5\textwidth}
\centering
\begin{tabular}{|c||r|r|r|r|}
\hline
\multicolumn{5}{|c|}{\boldmath $T_4$ \unboldmath} \\
\hline
         \boldmath $l=0$ \unboldmath & 1 & 0 & - & - \\
         
         \boldmath $l=1$ \unboldmath & -0.064 & 0.317 & 0 & - \\
         
         \boldmath $l=2$ \unboldmath & 0.103 & -0.136 & -0.004 & 0 \\
         
         \boldmath $l=3$ \unboldmath & 0.006 & -0.007 & 0.001&$\mathcal{O}$ \\
         
         \boldmath $l=4$ \unboldmath & 0.0003 & -0.001 &$\mathcal{O}$& $\mathcal{O}$ \\
         
         \boldmath $l=5$ \unboldmath & 0.00001 & $\mathcal{O}$ &$\mathcal{O}$&$\mathcal{O}$ \\
         
         \hline
\end{tabular}

\end{subtable}
\begin{subtable}{.5\textwidth}
\centering
\begin{tabular}{|c||r|r|r|r|}
\hline
\multicolumn{5}{|c|}{\boldmath $T_5$ \unboldmath} \\
\hline
         \boldmath $l=0$ \unboldmath & 0 & 0 & - & - \\
         
         \boldmath $l=1$ \unboldmath & 0.026 & 0 & 0 & - \\
         
         \boldmath $l=2$ \unboldmath & 0.003 & 0.035 & 0 & 0 \\
         
         \boldmath $l=3$ \unboldmath & 0.0003 & 0.001 & 0.001&$\mathcal{O}$ \\
         
         \boldmath $l=4$ \unboldmath & 0.00002 & 0.0002 &$\mathcal{O}$& $\mathcal{O}$ \\
         
         \boldmath $l=5$ \unboldmath & 0 & $\mathcal{O}$ &$\mathcal{O}$&$\mathcal{O}$ \\
         
         \hline
\end{tabular}

\end{subtable}

\caption{Numerical values for the coefficients $b_n^{(i,n+l)}$ of the polynomials $p_l^{(i)}(t)$ for the scalar functions $T_i$ up to $\mathcal{O}(1/m_b^5)$. For the values of the HQE parameters, the LLSA approximation is employed and we use $\hqe= 0.5$ GeV. The unknown coefficients of $\mathcal{O}(1/m_b^6)$ or higher are denoted by $\mathcal{O}$. }
\label{tab:b_coeff}
\end{table}

Note that our definition \eqref{pdef} of the coefficients $b_n^{(i,n+l)}$ already takes into account the 
growth factor $(2 l)!$. In case the factorial growth would be visible already 
in the terms up to $1/m_b^5$, the entries in Table~\ref{tab:b_coeff} should all be roughly of the same order. 
However, the picture we observe is not very conclusive, indicating that the factorial growth of the coefficients 
sets in only at even higher orders. Nevertheless, looking at 
Table~\ref{tab:b_coeff}, we see we can divide the five scalar functions into three groups when modelling 
$p_l^{(i)}(t)$: 
\begin{itemize}
    \item[(A)] $T_1$ and $T_4$: we assume all coefficients $b_n^{(i,n+l)}$ to be of the same order, i.e.\ equal to 1, except for the coefficients for terms independent of $t$ which vanish, i.e.\ $b_{l+1}^{(i,2l+1)}=0$. We therefore model the polynomials as  \begin{align}p^{(1,4)}_l(t)&= t^{l+1}+t^l+\ldots+t = \sum\limits_{m=1}^{l+1} t^m =\frac{t-t^{l+2}}{1-t}\ .\end{align}
    \item[(B)] $T_2$ and $T_3$: we assume all coefficients $b_n^{(i,n+l)}$ to be of the same order, i.e.\ equal to 1, except for vanishing coefficients for terms with the highest power in $t$ for each polynomial $p_l^{(i)}(t)$, i.e.\ $b_{0}^{(i,l)}=0$. We therefore model the polynomials as  \begin{align}p^{(2,3)}_l(t)&= t^l+\ldots +t+1 = \sum\limits_{m=0}^{l} t^m=\frac{1-t^{l+1}}{1-t}\ .\end{align}
    \item[(C)] $T_5$: we assume again all coefficients $b_n^{(5,n+l)}$ to be equal to 1, except for vanishing coefficients for terms independent of $t$ and terms linear in $t$, i.e.\ $b_{l+1}^{(5,2l+1)}=b_{l}^{(5,2l)}=0$. We therefore model the polynomial as \begin{align}
    p_0^{(5)}(t)&= 0\ ,\quad\quad
    p^{(5)}_{l\geq1}(t)= t^{l+1}+\ldots +t^2 =\sum\limits_{m=2}^{l+1} t^m\ ,\nonumber\\
    &\Rightarrow p^{(5)}_{l\geq0}(t)=\frac{t^2-t^{l+2}}{1-t}\ .
     \end{align}
\end{itemize}
\clearpage
Fixing the dependence of the $T_i$ on $t$ in this way, we can now proceed in studying the behaviour 
of \eqref{eq:Fakt} which we now consider to be the factorially growing contribution $T_i$ 
of the 
asymptotic series of the $T_i$, and we obtain for these terms
\begin{align}
    T_{1,4}(t,\lambda^2)&=\frac{1}{\hqe}\frac{t\lambda^2}{1-t}\left(F_1(\lambda)-tF_2(\lambda)\right)\ ,\nonumber\\
    T_{2,3}(t,\lambda^2)&=\frac{1}{\hqe}\frac{\lambda^2}{1-t}\left(F_1(\lambda)-tF_2(\lambda)\right)\ ,\nonumber\\
    T_{5}(t,\lambda^2)&=\frac{1}{\hqe}\frac{t^2\lambda^2}{1-t}\left(F_1(\lambda)-F_2(\lambda)\right)\ ,
\end{align}
where $\lambda=1/r$ and the $F_i$ correspond to the (formal) expressions
\begin{align}
    F_1(\lambda)&=\sum_{l=0}^\infty (2l)!(\lambda^2)^l\ ,\nonumber\\
    F_2(\lambda)&=\sum_{l=0}^\infty (2l)!(t\lambda^2)^l\ .
\end{align} 

Making use of the procedure and definition of DV in Sec.~\ref{sec:def_dv}, we use the 
Borel transform to define the ambiguities in the transformation of the asymptotic series $F_1(\lambda)$ and $F_2(\lambda)$, similar to equation \eqref{eq:F_DV}. Inserting the results for $\Delta_{\rm{DV}} F_1(\lambda)$ and $\Delta_{\rm{DV}} F_2(\lambda)$ into the expressions for $T_i$, we identify the outcome with 
the DV contributions to the $T_i$. However, in the decay rate $T_{\mu\nu}$ does not enter but rather the hadronic tensor $W_{\mu\nu}$. Using the same Lorentz decomposition for $W_{\mu\nu}$ as for $T_{\mu\nu}$ allows for the structure functions $W_i$ to be obtained using the optical theorem $W_i=-\frac{1}{\pi}\text{Im}\ T_i$. Applying this to our duality-violating terms we find  
    \begin{align} \label{eq:T14}
        \hat\Delta_{\rm{DV}}W_{1,4}(vQ,Q^2)=-\frac{1}{\pi}\hat\Delta_{\rm{DV}} {\rm Im} &\left[T_{1,4}(vQ,Q^2)\right] = &\\
\frac{1}{\hqe-vQ}&\frac{vQ}{\sqrt{Q^2}}\left(\sin{\left(\frac{\sqrt{Q^2}}{\hqe}\right)}-\sqrt{\frac{vQ}{\hqe}}\sin{\left(\frac{1}{\sqrt{\hqe}}\sqrt{\frac{Q^2}{vQ}}\right)}\right)
   \nonumber	
    \end{align}
\begin{align} \label{eq:T23}
        \hat\Delta_{\rm{DV}}W_{2,3}(vQ,Q^2)=-\frac{1}{\pi}\hat\Delta_{\rm{DV}} {\rm Im} &\left[T_{2,3}(vQ,Q^2)\right] = &\\
\frac{1}{\hqe-vQ}&\frac{\hqe}{\sqrt{Q^2}}\left(\sin{\left(\frac{\sqrt{Q^2}}{\hqe}\right)}-\sqrt{\frac{vQ}{\hqe}}\sin{\left(\frac{1}{\sqrt{\hqe}}\sqrt{\frac{Q^2}{vQ}}\right)}\right)
   \nonumber
    \end{align}
        \begin{align} \label{eq:T5}
       \hat\Delta_{\rm{DV}}W_{5}(vQ,Q^2)=-\frac{1}{\pi}\hat\Delta_{\rm{DV}} {\rm Im} &\left[T_{5}(vQ,Q^2)\right] = &\\
\frac{1}{\hqe-vQ}&\frac{(vQ)^2}{\hqe\sqrt{Q^2}}\left(\sin{\left(\frac{\sqrt{Q^2}}{\hqe}\right)}-\sqrt{\frac{\hqe}{vQ}}\sin{\left(\frac{1}{\sqrt{\hqe}}\sqrt{\frac{Q^2}{vQ}}\right)}\right) .
   \nonumber
    \end{align}
As per our schematic expectation, the contribution of DV to the structure functions is a sinusoidal function. The splitting of the five scalar functions into three groups has led to three slightly different behaviours in the amplitudes. Due to the fact that, in our model, all $T_i$ are functions of the same asymptotic series, the arguments of the sinusoidal functions are the same for all $\hat\Delta_{\rm{DV}}W_{i}(vQ,Q^2)$. Finally, it is clear that the choice of $\hqe$ will have an impact on the resulting DV. In the following section, we will discuss how the choice of $\hqe$ affects observables in inclusive decays.

\section{The QHDV model}\label{sec:DVModels}
 The triple differential rate, for the Lorentz decomposition of $T_{\mu\nu}$ and thus equivalently $W_{\mu\nu}$ in \eqref{eq:T_decomp}, is given by 
\begin{align}
    \frac{\text{d}^3\Gamma}{\text{d}\hat{q}^2\text{d}s\text{d}y}=&\ 48m_b\Gamma_0\Bigg[\frac{2y s-y^2-2\hat{q}^2+y\hat{q}^2}{1-s}W_1+\hat{q}^2W_2+\frac{1}{2}\left(2y s-y^2-\hat{q}^2\right)W_3\nonumber\\
    &+\frac{2y s-y^2-\hat{q}^2}{1-s}W_4+\frac{2y s-y^2-\hat{q}^2}{2(1-s)^2}W_5\Bigg]\theta\left(\hat{q}^2\right)\theta\left(2y s-y^2-\hat{q}^2\right)\ , \nonumber\\
    \Gamma_0=& \frac{G_F^2|V_{cb}|^2m_b^5}{192\pi^3}\, .\label{eq:triple}
\end{align}
Here we have introduced dimensionless variables \begin{align}
    \hat{q}^2=\frac{q^2}{m_b^2}\ ,\quad s= \frac{v\cdot q}{m_b}\ ,\quad y= \frac{2E_\ell}{m_b}\ ,
\end{align}
with  $E_{\ell}$ the lepton  energy and $q^2$ the leptonic invariant mass of the $B(p_B)\to X_c(p_c)\ell(p_\ell)\bar{\nu}(p_\nu)$ decay. 

The duality violating contributions to the hadronic tensor modelled in \eqref{eq:T14}, \eqref{eq:T23} and \eqref{eq:T5} enter the kinematic variables of inclusive decays together with the OPE contribution to $W_i$. However, from the above 
construction, we do not have an absolute normalisation of the DV terms compared to the contributions 
of the OPE, since we only can infer the dependence of the DV terms on the kinematic variables. 
Thus we multiply the DV terms for the $W_i$ by a normalization constant $N_i$
\begin{align}
    W_i\to W_i^{(\rm{OPE})}+N_i \hat\Delta_{\rm{DV}}W_i(s,\hat q^2, \hqe)\ .
\end{align}
The dimensionless normalization constant $N_i$ determines the strength of the quark-hadron duality violations, which should be determined from the experimental data. In principle, $N_i$ can be different for each $W_i$ contribution. However, we assume $N_i= N$ for all $i=1,\ldots, 5$. In addition, we normalize the QHDV contribution through
\begin{equation}
    N_i = N =\frac{\Gamma_{\rm{P}}}{\Gamma_{\rm{DV}}}\,\,\mathcal{C}_{\rm{DV}}\ ,
\end{equation}
where $\Gamma_{\rm P}$ is the partonic rate 
\begin{equation}
    \Gamma_{\rm{P}}=\Gamma_0(1-8\rho+8\rho^3-\rho^4-12\rho^2\log{\rho})\ ,\quad\quad  \rho\equiv \frac{m_c^2}{m_b^2}\ , 
\end{equation}
and $\Gamma_{\rm DV}$ is the unnormalized DV contribution found by integrating the differential rate in \eqref{eq:triple} with the replacement $W_i\to \hat\Delta_{\rm{DV}} W_i(s,\hat q^2,\hqe)$.  Note that this normalisation depends on the choice of $\hqe$. Taking $\hqe=0.5$ GeV, we find $\Gamma_{\rm P}/\Gamma_{\rm DV}=0.2508$. Specifically, the normalization is chosen in such a way that 
\begin{equation}\label{eq:totrat}
            \frac{\Gamma}{\Gamma_0} = 0.657 + 0.657\,\, \mathcal{C}_{\rm DV}  -0.025|_{m_b^2}  -0.026|_{m_b^3} + 0.0003|_{m_b^4} + 0.007|_{m_b^5}  \ ,\\ 
\end{equation}
i.e. the partonic contribution and the DV contribution are of the same size for $\mathcal{C}_{\rm{DV}}=1$ (and $\hqe=0.5$ GeV). 

Finally, our model for QHDVs thus only depends on $\mathcal{C}_{\rm{DV}}$ and $\Lambda_{\rm HQE}$. 

\subsection{Differential spectra}

\begin{figure}[h!]
    \centering
    \begin{subfigure}[t]{0.48\textwidth}
    \centering
    \includegraphics[width=0.9\textwidth]{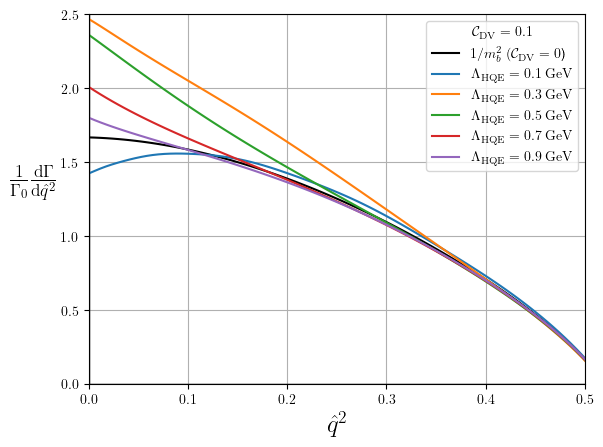}
    \end{subfigure}%
    \begin{subfigure}[t]{0.48\textwidth}
    \centering
    \includegraphics[width=0.9\textwidth]{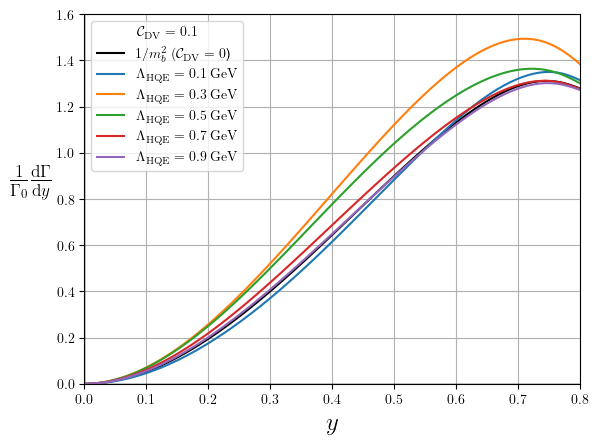}
    \end{subfigure}%

    \caption{Dependence of the differential rates (up to $\mathcal{O}(1/m_b^2)$ including the QHDV contributions) on different values of $\Lambda_{\rm HQE}$ for $\mathcal{C}_{\rm{DV}}=0.1$. The left and right plots show the differential rate with respect to $\hat q^2=q^2/m_b^2$ and $y=2E_\ell/m_b$ respectively. }
    \label{fig:diff_spectra} 
 \end{figure}

Integrating the differential rate in \eqref{eq:triple} with the replacement $W_i\to W_i^{(\rm{OPE})}+N\hat\Delta_{\rm{DV}}W_i(s,\hat q^2,\hqe)$ allows us to determine the differential $\text{d}\Gamma/\text{d}\hat{q}^2$ and $\text{d}\Gamma/\text{d}y$. In Fig.~\ref{fig:diff_spectra}, we show these spectra for possible different values of $\hqe$. To do so, we keep the normalisation, defined for fixed  $\hqe=0.5$ GeV, constant at $N= 2.508\ \mathcal{C}_{\rm{DV}}$ and take $\mathcal{C}_{\rm DV}=0.1$. Moreover, we only show the OPE result up to $\mathcal{O}(1/m_b^2)$, since at higher orders (derivatives of) delta-functions will occur, which would cause divergences in the differential rate. The inputs are given in Appendix~\ref{app:inputs}. We stress that due to the fixed normalization the effect of QHDV also depends on the choice of $\hqe$. In addition, we see that the value of $\hqe$ slightly affects the shape of the differential spectra. 

We do not clearly see the expected oscillation. This is because in these examples the period of the QHDV function is too big for the oscillation to be visible in the kinetically allowed regions of $\hat{q}^2$ and $y$. Therefore, we do not see the characteristic ``wiggle'' around the OPE result. As said, our model assumption is that both the strength $\mathcal{C}_{\rm DV}$ and the scale $\hqe$ are free parameters. Nevertheless, the setup of the QHDV from the HQE suggests a typical scale for the duality violation of the order of $\hqe = 0.5$ GeV and motivates the range for $\hqe$ used in Fig.~\ref{fig:diff_spectra}. 

\subsection{Comparison to Instanton-Induced Duality Violation  }
In the context of the discussion about possible duality violation in the HQE,
it has been noticed that instantons can induce terms which do not allow for an 
expansion in inverse powers of the heavy-quark mass \cite{Chay:1994si,Falk:1995yc,Chibisov:1996wf}. 
The calculations employ the propagator of the final state quarks in a background field of an instanton, 
which introduces a dimensionful parameter $\omega$ corresponding to the size of the instanton.
This parameter corresponds to the scale $\hqe$ appearing in our model, which -- by our construction -- 
is of the order of $\qcd$. However, the conclusion of \cite{Chay:1994si,Falk:1995yc,Chibisov:1996wf}
is that the instanton contribution suffers from a strong suppression, corresponding to 
\begin{equation}
N_i \sim \frac{1}{m_b^{6 \cdots 8}} \, , 
\end{equation}
depending on the observable under consideration. The overall conclusion is that instanton-induced 
duality violation is irrelevant at the current level of precision.  

However, it is worthwhile to notice that the instanton-induced 
duality violation in differential rates is proportional to \cite{Chibisov:1996wf}
\begin{equation}
\sin \left(2 \omega \sqrt{(m_b v - q)^2} \right)\ ,
\end{equation} 
which indicates that the scale $\hqe$ appearing in the duality-violating terms is not 
necessarily of order $\qcd$. 

The model for duality violation we are proposing here is based on an analysis of the HQE up to $1/m_b^5$ contributions. As we pointed out above, there is no clear indication that the asymptotic behaviour of the HQE is visible already at such low order in the expansion. In fact, if the asymptotic behaviour (as suggested by our numerical analysis) sets in only at higher orders, the scale $\hqe$ appearing in (\ref{eq:T14}, \ref{eq:T23}, \ref{eq:T5}) could be replaced by a generic scale $\dv$, which in principle can be independent of $\qcd$ and/or $\hqe$.

\begin{figure}
    \centering
      \includegraphics[width=0.48\textwidth]{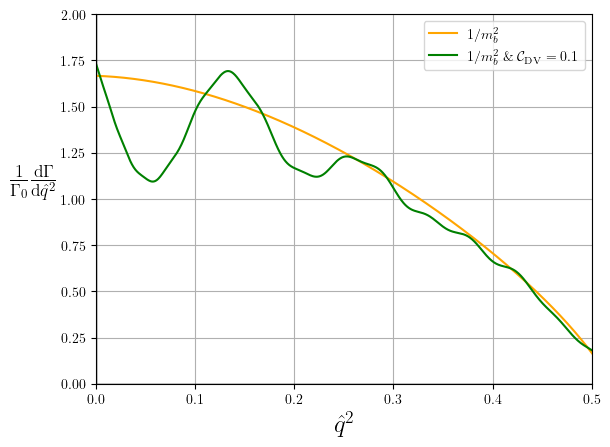}
    \caption{The DV contribution to the differential $\hat q^2$ spectrum with $\mathcal{C}_{\rm{DV}}=0.1$ and $\dv=10^{-4}$ GeV. The yellow line represents the OPE contribution to $1/m_b^2$ and the green line also includes the DV contribution. }
    \label{fig:wiggle}
\end{figure}

This motivates us to interpret our QHDV model as having two free parameters, namely $\cdv$ and to replace $\hqe$ with the more generic scale $\dv$. These parameters are now completely free fit parameters to be constrained by experimental data. It is then interesting to consider much smaller values for $\dv$. In Fig.~\ref{fig:wiggle}, we show the differential $q^2$ spectra as in Fig.~\ref{fig:diff_spectra} but now with $\dv=10^{-4}$ GeV. At small scales like this, we can see that our model shows the characteristic oscillatory behaviour around the OPE result.

\section{QHDV in kinematical moments}\label{sec:kin_mom}
In order to probe the effect of possible QHDV contribution, the parameters $\mathcal{C}_{\rm DV}$ and $\dv$ should be constrained by data. In practice, we cannot use the differential spectra, due to the singular functions appearing when including higher-order terms in the HQE. Therefore, we need to consider integrated observables. In the following, we consider moments of both the integrated $q^2$ and lepton energy $E_\ell$ differential spectrum\footnote{For simplicity, we do not consider here $M_X^2$, which only differs from the charm mass at order $\alpha_s$. As such $\alpha_s$ corrections are important, and it was found that $\alpha_s^3$ corrections are particularly large for $M_X^2$ moments \cite{Fael:2022frj}.}. We define these moments as 
\begin{equation} 
 Q_n(q^2_{\rm cut})  \equiv \frac{1}{\Gamma_0} \int_{q^2_{\rm cut}} \text{d}q^2 \,  (q^2)^n \frac{\text{d} \Gamma}{\text{d} q^2}\ ,
\end{equation}
and
\begin{equation} 
L_n(E_\ell^{\rm cut}) \equiv \frac{1}{\Gamma_0} \int_{E_\ell^{\rm cut}} \text{d}E_\ell \,  E_\ell^n \frac{\text{d} \Gamma}{\text{d} E_\ell}\ ,
\end{equation}
where we include a kinematical constraint on $q^2$ and $E_\ell$. %

For simplicity, we consider here the QHDV effects on the normalized moments, defined through\footnote{Often centralized moments are considered. To simplify the discussion and show the effect of QHDV, we consider here only normalized moments.}
\begin{equation}
\bar{q}_n \equiv \langle (q^2)^n \rangle_{ q^2\geq q^2_{\rm cut} }   \equiv   \frac{Q_n(q^2_{\rm cut})}{Q_0(q^2_{\rm cut})}\ ,  \quad\quad \bar{\ell}_n \equiv \langle E_\ell^n \rangle_{E_\ell\geq E_\ell^{\rm cut}}   \equiv   \frac{L_n(E_\ell^{\rm cut})}{L_0(E_\ell^{\rm cut})} \ .
\end{equation}
We note that due to this normalization, it is customary to re-expand the ratios both in $\alpha_s$ and $1/m_b$ terms. Here we consider QHDV terms to be small,and therefore also re-expand in $\cdv$. When re-expanding, we thus neglect all HQE elements that multiply duality-violating terms, i.e. we neglect $\mathcal{C}_{\rm DV}/m_b$-terms. %
In Table~\ref{tab:contris}, we show the relative contribution of the power corrections and the QHDV term, where for the latter we assume $\mathcal{C}_{\rm DV} = 1$. In addition, we use as our default value $\hqe=0.5$ GeV. Since the DV contribution comes in linearly after re-expanding in $1/m_b$ and $\cdv$, its effect can easily be gauged.
We use the numerical inputs given in Appendix~\ref{app:inputs}. For simplicity, we have assumed no lepton energy nor $q^2$ constraints. We note that the size of the power corrections stems from assuming LLSA values for all $1/m_b^{n}$ terms and is just merely an indication. 

We recall that our normalization is such that $\cdv=1$ implies that QHDV effects are equal to the partonic contribution of the total rate \eqref{eq:totrat}. For the moment, we see from Table~\ref{tab:contris} that QHDV contributions are sizeable for $\cdv=1$, especially for the $q^2$ moments. Comparing to the contribution of the power corrections shows that if $\cdv \simeq 0.01$ the QHDV contribution is of the same order as the $1/m_b^5$ contribution.

\begin{figure}[t!]
    \centering
    \includegraphics[width=0.40\textwidth]{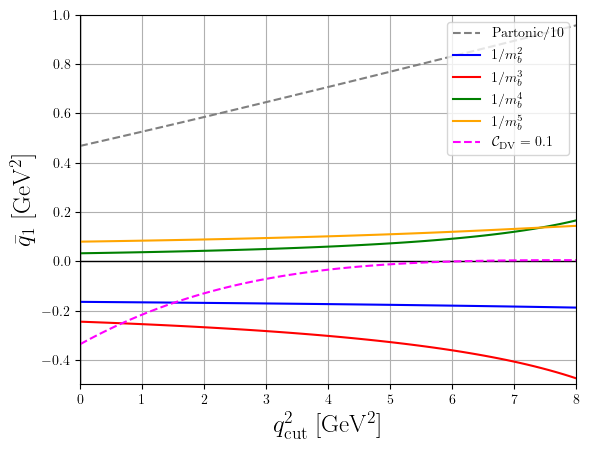}
    \includegraphics[width=0.40\textwidth]{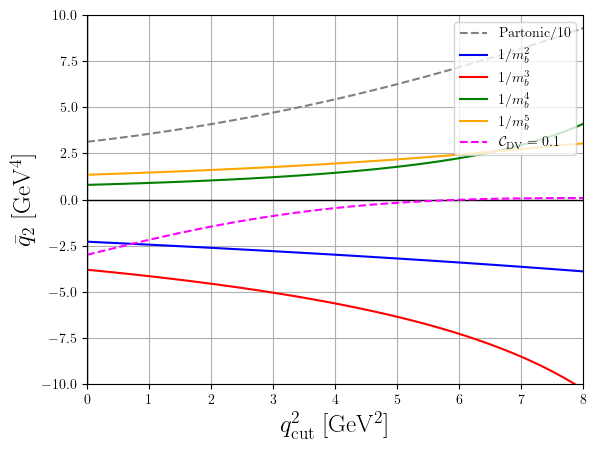}
    \includegraphics[width=0.40\textwidth]{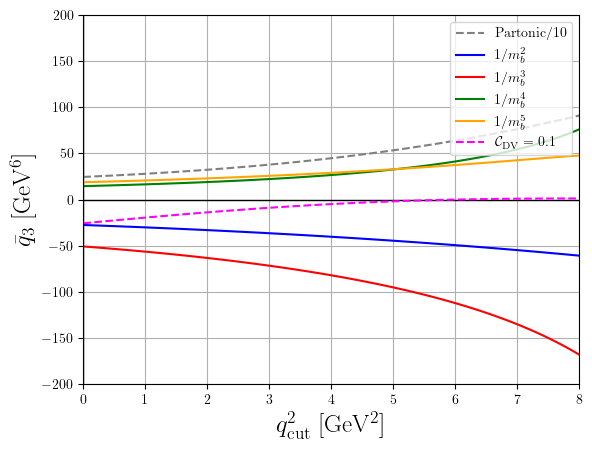}
    \includegraphics[width=0.40\textwidth]{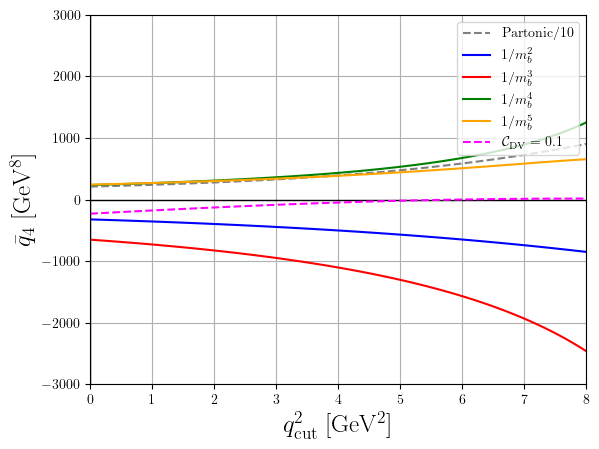}
     \caption{The $\bar{q}_n$ moments as a function of $q^2_{\rm{cut}}$, split into the different partonic, $1/m_b$, and DV contributions. The different coloured solid lines indicate the contribution from the different power corrections. The dashed grey and magenta lines indicate the partonic and DV contributions, respectively. Note that the partonic contribution is scaled down by a factor of 10.     }
    \label{fig:q2moments_DV_split}
\end{figure}

\begin{figure}[t!]
    \centering  
    \includegraphics[width=0.40\textwidth]{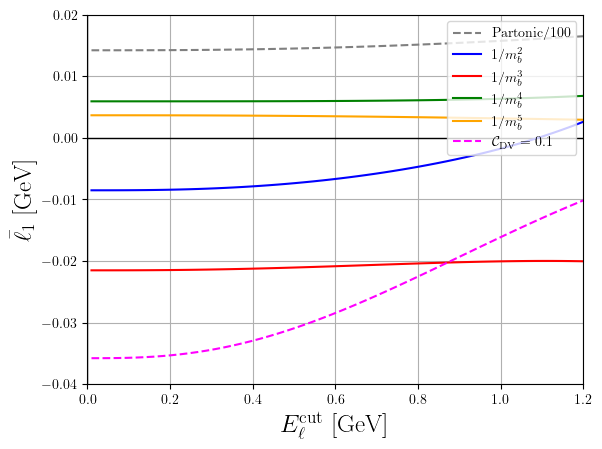}
    \includegraphics[width=0.40\textwidth]{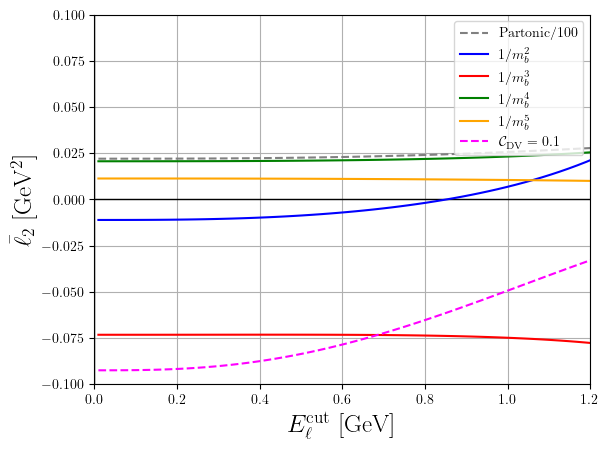}
    \includegraphics[width=0.40\textwidth]{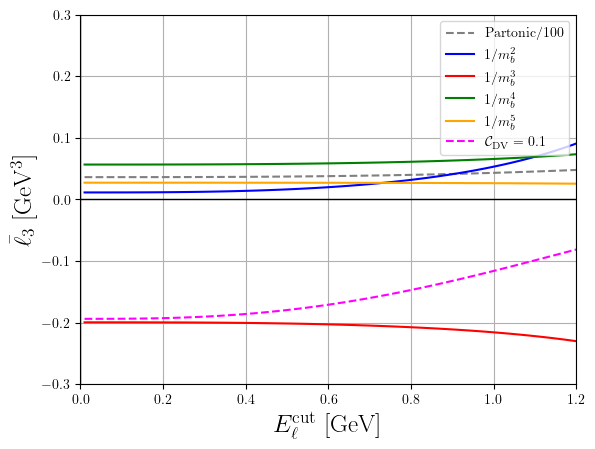}
    \includegraphics[width=0.40\textwidth]{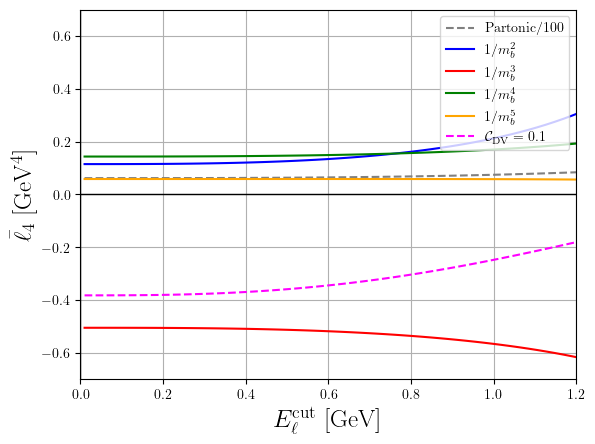}
    \caption{The $\bar{\ell}_n$ moments as a function of $E_\ell^{\rm{cut}}$, split into the different partonic, $1/m_b$, and DV contributions. The different coloured solid lines indicate the contribution from the different power corrections. The dashed grey and magenta lines indicate the partonic and DV contributions, respectively. Note that the partonic contribution is scaled down by a factor of 100.}
    \label{fig:Elmoments_DV_split}
\end{figure}

In Figs.~\ref{fig:q2moments_DV_split} and \ref{fig:Elmoments_DV_split}, we show the dependence of the $\bar{q}_i$ and $\bar{l}_i$ moments on their kinematical cuts for fixed $\cdv=0.1$ and $\hqe=0.5$ GeV. The total has been split into different contributions from QHDV and power corrections. Note that the partonic results for $\bar{q}_n$ and $\bar{\ell}_n$ are divided by a factor of 10 and 100, respectively. We can see from Fig.~\ref{fig:q2moments_DV_split} that the QHDV contribution is most significant when $q^2_{cut}$ approaches zero. This is a direct result of the QHDV differential rate $\text{d}\Gamma_{\rm{DV}}/\text{d}\hat{q}^2$ being large at small values of $\hat{q}^2$ as can be seen in Fig.~\ref{fig:diff_spectra}. On the other hand, the power corrections actually become larger for higher $\hat{q}_{cut}^2$. 
In Fig.~\ref{fig:Elmoments_DV_split} we can see that, similar to the case of $q^2$ moments, the QHDV contribution to $\bar{\ell}_n$ is largest at small $E^{cut}_{\ell}$. Finally, we also considered the effect on the forward-backward asymmetry $\mathcal{A}_{\rm FB}$ as a function of $q^2_{\rm cut}$ \cite{Herren:2022spb}. We find that the QHDV contributions have the same dependence on the cut as the power corrections.

\renewcommand{\arraystretch}{1.2}
\begin{table}[t!]
    \centering
    \begin{tabular}{|c||c|c|c|c|c|c|}
    \hline
    \textbf{Moment} & \textbf{Partonic} & \textbf{QHDV} & \boldmath ${1/m_b^2} $ &\boldmath ${1/m_b^3} $ & \boldmath ${1/m_b^4 }$& \boldmath ${1/m_b^5 }$\\ 
    \hline   \hline
        \boldmath ${\bar{\ell}_1}$ $(\text{GeV})^{\phantom{1}}$ & 1.4 & -0.36 & -0.009 & -0.022 & 0.006& 0.004 \\   \hline
        \boldmath ${\bar{\ell}_2}$ $(\text{GeV}^2)$& 2.2 & -0.93 & -0.011 &-0.074   & 0.021 &0.011\\   \hline
        \boldmath ${\bar{\ell}_3}$  $(\text{GeV}^3)$& 3.6 &-1.94 & -0.011 & -0.201 & 0.056 & 0.027\\   \hline
         \boldmath ${\bar{\ell}_4}$  $(\text{GeV}^4)$ & 6.1 &-3.83 & 0.114 & -0.508 & 0.143 & 0.058\\   \hline
         \boldmath ${\bar{q}_1}$  $(\text{GeV}^2)$ & 4.7 & -3.4 & -0.165 & -0.245 &0.032  &0.079 \\   \hline
         \boldmath ${\bar{q}_2}$  $(\text{GeV}^4)$& 31.3 & -29.9 & -2.276  &-3.793  & 0.799 & 1.347  \\   \hline
            \boldmath ${\bar{q}_3}$  $(\text{GeV}^6)$& 245.9 & -256.1   & -27.44  & -50.66  &  14.61 & 18.83 \\ \hline   
               \boldmath ${\bar{q}_4}$  $(\text{GeV}^8)$&  2116 & -2278 & -320.7 & -650.0 & 237.17 & 243.9 \\   \hline
    \end{tabular}
    \caption{Normalized moments and their relative dependence of the QHDV contribution with respect to the partonic and power corrections. Here we have put $\mathcal{C}_{\rm DV} = 1$ and $\hqe=0.5$ GeV. All coefficients are in GeV to the appropriate power.} 
    \label{tab:contris}
\end{table}
\renewcommand{\arraystretch}{1}

\clearpage

\section{Quantifying Duality Violations from Data} \label{sec:odv}

As pointed out in the introduction, there is currently no indication of large duality violations, 
since the known terms in the $1/m_b$ expansion yield a consistent picture. This indicates that we 
are either still far away from the order in the HQE, where the asymptotic behaviour sets in, or that
the duality violations are overall small or even absent. 

In principle, our model for QHDV could be included in a global fit to the available moments to determine $\mathcal{C}_{\rm DV}$ and the HQE parameters at the same time. However, since we do not know the size of the 
higher-order terms of the HQE series, it may be hard to disentangle a small duality-violating effects 
from the unknown higher orders in $1/m_b$.

To this end, it is interesting to construct observables $O_{\rm DV}^{(k)}$, which do not 
have any contributions of lower orders in the HQE, i.e. 
\begin{equation} \label{Obsk} 
O_{\rm DV}^{(k)}\sim \hqe^{k+1}/m_n^{k+1} \ .
\end{equation}
This can always be achieved by linear combinations of observables ${\cal K}_j$, for which an HQE 
can be set up. In order to find an observable which satisfies \eqref{Obsk}, one needs to use 
$l+2$ observables ${\cal K}_j$, where $l$ is the number of HQE parameters appearing up to order 
$\hqe^k/m_b^k$. The condition that in the linear combination of ${\cal K}_j$ the coefficients 
of all HQE parameters vanish, defines a linear system of equations, which can be solved to find these
coefficients. 

As an example, we consider the $q^2$ moments\footnote{A similar analysis can be set up for the lepton energy moments}, for which we write the expansion up to order $\hqe^3/m_b^3$ 
\begin{equation}
    \bar{q}_i = C_i^{(0)} + \frac{\mu_G^2}{m_b^2} C_i^{(2)} +\frac{\tilde{\rho}_D^3}{m_b^3}C^{(3)}_i+ R_i \ ,
\end{equation}

where the $R_i$ are the residual terms of higher order in the HQE and/or duality-violating terms. In addition, in principle the $C_i^{(j)}$ depend on $m_c, m_b$ and $\alpha_s$. Note that $\mu_{\pi}^2$ and $\rho_{LS}^3$ do not enter because of the reparametrisation invariance of the $q^2$ moments. We now define a linear combination of the moments, which consists of the residual terms only, which in the HQE is of order $\hqe^4/m_b^4$ 
\begin{equation}
O_{\rm DV}^{(3)} = \xi_1 \frac{\bar{q}_1}{m_b^2} + \xi_2 \frac{\bar{q}_2}{m_b^4} + \xi_3 \frac{\bar{q}_3}{m_b^6} +\xi_4 \frac{\bar{q}_4}{m_b^8}\ ,
\end{equation} 
where the $\xi_i$ are determined in terms of the $C_i^{(j)}$ and which depend on the kinematic cut on the moments. This amounts to solving the equations
\begin{align}
    \xi_1 \frac{C_1^{(n)}}{m_b^2}+\xi_2 \frac{C_2^{(n)}}{m_b^4}+\xi_3 \frac{C_3^{(n)}}{m_b^6}+\xi_4 \frac{C_4^{(n)}}{m_b^8}=0\quad\quad (n=0,2,3)\ ,
    \label{eq:xis}
\end{align}
for $\xi_{2,3,4}$
and leaving $\xi_1$ as an arbitrary normalization constant. The extension of this idea to higher orders is
evident. Note that since the measurements are available at different kinematical cuts, we can solve \eqref{eq:xis} at for each kinematic cut. This results in multiple distinct, but correlated, observables. 

By solving equation \eqref{eq:xis}, we obtain $\xi_{2,3,4}$ as a function $q^2_{\rm cut}$ and $\xi_1$. Choosing $\xi_1 =1$, we can express $O_{\rm DV}^{(3)}$ in terms of $R_i$. Using the HQE expressions up to $1/m_b^5$ for $R_i$ and including duality violations, we find e.g.
\begin{align}
O_{\rm DV}^{(3)}&= (5.182\,\, \mathcal{C}_{\rm DV}  -0.546|_{m_b^4} + 0.519|_{m_b^5} ) \times 10^{-3}  \quad\quad (q^2_{\rm cut}=3.0\ \text{GeV}^2)\ ,\nonumber  \\ 
O_{\rm DV}^{(3)}&= (2.166\,\, \mathcal{C}_{\rm DV}  -0.494|_{m_b^4} + 0.499|_{m_b^5} ) \times 10^{-3} \quad\quad (q^2_{\rm cut}=3.0\ \text{GeV}^2)\ ,\nonumber  \\  
O_{\rm DV}^{(3)}&= (0.751\,\, \mathcal{C}_{\rm DV}  -0.447|_{m_b^4} + 0.487|_{m_b^5} ) \times 10^{-3}  \quad\quad (q^2_{\rm cut}=3.0\ \text{GeV}^2)\ , 
\label{eq:ODV_theory}
\end{align}
where, as before, we use the numerical estimates for the HQE parameters listed in Appendix~\ref{app:inputs}. In \eqref{eq:ODV_theory}, we give $O_{\rm DV}^{(3)}$ for three different cuts, similarly we can calculate the theory expression for other cuts or for other moments. We see that the power corrections at $1/m_b^4$ and $1/m_b^5$ almost perfectly cancel in these observables. Within the LLSA estimates for the $1/m_b^{4,5}$ terms, we thus claim that $O^{(3)}_{\rm DV}$ is very sensitive to duality violations and/or higher-order corrections in the HQE.   

\begin{figure}[t!]
    \centering
    \includegraphics[width=0.62\textwidth]{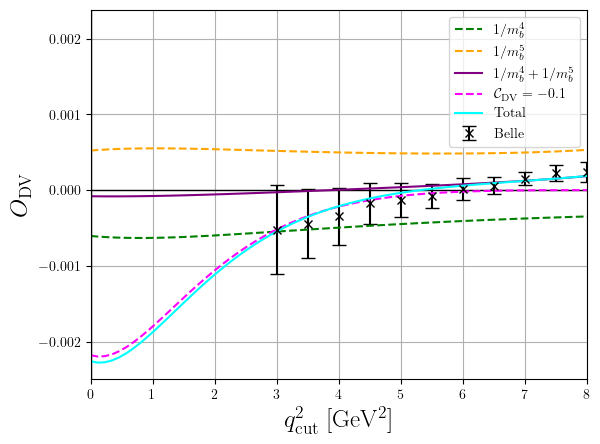}
    \caption{Theoretical predictions for the different contributions to the $O_{\rm{DV}}^{(3)}$. The data points are constructed from $q^2$ moments measured by \cite{Belle:2021idw}.}
    \label{fig:ODV_theory_data}
\end{figure}

The $\bar{q}_i$ moments have been measured by Belle \cite{Belle:2021idw} and Belle II 
\cite{Belle-II:2022evt} as a function of $q^2_{\rm cut}$ starting at $3$ GeV$^2$. 
These data can be used to obtain an experimental value for $O^{(3)}_{\rm DV}$, which can be directly 
related to higher order terms, or likewise to duality violation using \eqref{eq:ODV_theory}. We proceed by comparing this experimental value 
to the model expressions for duality violation and determine the $\mathcal{C}_{\rm DV}$. We use only the $q^2$ moments from Belle \cite{Belle:2021idw} for the electron channel. Taking the correlations between the $q^2$ moments into account, we can construct $O^{(3)}_{\rm{DV}}$ for each $q^2$-cut using also input values for $m_b$ and $m_c$ entering through the $C_i^{(j)}$. In principle, the $O_{\rm{DV}}^{(3)}$ also have $\alpha_s$ corrections, but for this first study we do not take those into account. The  $O_{\rm{DV}}^{(3)}$ constructed from data are given in Fig.~\ref{fig:ODV_theory_data} in black. We find that $O_{\rm{DV}}^{(3)}$ is consistent with zero within uncertainties. Here we also show the theoretical prediction of $O_{\rm{DV}}^{(3)}$. We again observe the cancellation between the $1/m_b^4$ and $1/m_b^5$ terms. Here, we show as well the QHDV contribution for $\mathcal{C}_{\rm{DV}} = -0.1$, inspired by the apparent trend in the experimental data. The effect of QHDV is largest at lower experimental cuts. 

Combining the experimental data from \cite{Belle:2021idw} with the the expression in \eqref{eq:ODV_theory}, we can determine $\cdv$ for each $\mathcal{O}_{\rm DV}^{(3)}(q^2_{\rm cut})$. As an example, we use $\mathcal{O}_{\rm DV}^{(3)}$ at $q^2_{\rm cut} =3,4$ and $5$ GeV$^2$, because these observables will impose the strongest constraint on QHDV. From these observables, we find
\begin{align}
\mathcal{C}_{\rm{DV}} &= -0.10\pm0.11 \quad\quad (q^2_{\rm cut}=3.0\ \text{GeV}^2)\ , \nonumber \\ 
\mathcal{C}_{\rm{DV}} &= -0.16\pm0.17 \quad\quad (q^2_{\rm cut}=4.0\ \text{GeV}^2)\ ,\nonumber \\ 
\mathcal{C}_{\rm{DV}} &= -0.30\pm0.30  \quad\quad (q^2_{\rm cut}=5.0\ \text{GeV}^2)\ ,
\label{eq:CDV_res}
\end{align}
respectively. The extractions at different kinematic cut show a consistent picture, and the above determinations could be combined to yield an even stronger constraint. Therefore, it would be interesting to do a full analysis of all available data in the future, including possibly lepton energy moments to further constrain $\cdv$. Our first simple data-driven study already shows, as expected, that the duality violation effects are consistent with zero. On the other hand, the uncertainties are still rather large.

Finally, we note that in order to obtain the constraints in \eqref{eq:CDV_res}, we assumed numerical values for the HQE parameters using the LLSA. However, even without this assumption a study of these new sensitive observables is useful. Comparing several independent observables $O_{\rm DV}^{(k)}$ constructed at different cuts and for different moments allows for data-driven insights into the higher-order $k+1$ terms of the HQE. If all the observables $O_{\rm DV}^{(k)}$ are decreasing according to their ``natural size'' $\hqe^k/m_b^k$ it is impossible (and also irrelevant) to disentangle effects from tiny duality violations from higher-order HQE terms. In this case, we would be still far away from the asymptotic regime, meaning that the HQE can be trusted at the precision level indicated by the natural power counting. On the other hand, if some or all of the observables $O_{\rm DV}^{(k)}$ turn out not to behave like $\hqe^k/m_b^k$, we would interpret this as the onset of asymptotic behaviour, i.e. as an indication of duality violation.

\section{Conclusions}\label{sec:concl}
The Heavy Quark Expansion -- in particular for the inclusive semileptonic $b \to c$ transition -- has been developed to 
impressive order in both the $\hqe / m_b$ as well as in the $\alpha_s$ expansion. Although there are 
convincing conjectures that the HQE eventually is an asymptotic series, the current state-of-the-art analyses do not show such an onset of diverging behaviour. 

At the same time, the inclusion of NNLO corrections to the moments and even N$^3$LO QCD corrections to the total rate, allows for e.g. an extraction of the CKM element $|V_{cb}|$ with percent-level theoretical  
uncertainty. When pushing this uncertainty down, the question on the behaviour of higher orders in $\hqe / m_b$ becomes relevant, and 
-- in particular -- if there is already an effect of a possible asymptotic behaviour visible at order $\hqe/m_b^{4,5}$. In order to quantify this, we have {\it assumed} that the HQE is asymptotic, meaning that 
at some order the coefficients exhibit a factorial growth, which we take as a definition of Quark-Hadron 
Duality violation. We proposed a model for these effects, based on the behaviour of the known 
HQE coefficients up to and including terms of order $(\hqe /m_b)^5$. Assuming that the series is still 
Borel summable, we compute (within our model) an explicit form of duality-violating terms by studying 
the resulting ambiguities.  

Applying this technique to inclusive $B \rightarrow X_c \ell \bar{\nu}$ decays, we quantified the effect of duality violation on the kinematic moments of the decay. In our model, the DV 
contribution depends only on two parameters: a hadronic scale $\dv$ and an overall coefficient $\mathcal{C}_{\rm{DV}}$, since we can only determine the shape of the
DV contribution but not its absolute size. 

In order to quantify this, we suggest to construct observables 
$O_{\rm DV}^{(k)}$, which only depend on DV and terms of order $(\hqe/m_b)^{k+1}$ or higher. A measurement of these observables will allow insight in the convergence of the HQE, while also shedding light 
on the size of the higher-order terms and on possible duality violations. 
As an example, we have constructed an observable $O_{\rm DV}^{(3)}$ from the $q^2$ moments. Using the “lowest-lying state saturation ansatz” (LLSA) the $1/m_b^4$ and $1/m_b^5$ contributions to $O_{\rm{DV}}^{(3)}$ are found to mostly cancel out, such that $O_{\rm DV}^{(3)}$ in fact only depends on DV and corrections of $(\hqe/m_b)^6$ or higher. 

Using the measured $q^2$ moments from \cite{Belle:2021idw}, we calculated the experimental $O_{\rm{DV}}^{(3)}$ at different $q^2$-cuts. We find that these observables are in agreement with zero within uncertainties. For low values of the $q^2_{\rm cut}$, we also determined $\cdv$ for the first time directly from the experimental data. We find, 
\begin{equation}
    \cdv = -0.1\pm 0.1 \ ,
\end{equation}
for the lowest $q^2_{\rm cut}$ and similar results for higher cuts. 

We emphasize that our results are fully consistent with a total absence of 
duality violation. However, this approach to control the HQE can be refined 
in many ways, e.g. by constructing observables for higher orders, also from 
different observables such as other kinematic moments, and by including QCD corrections. However, our main conclusion is that a determination of $|V_{cb}|$ with a theoretical uncertainty of about $1\%$ will not be obstructed by duality violation.  
However, with the methods provided here, one can test a possible limitation by duality violation when future higher precision determinations are attempted.

\section*{Acknowledgements}
This research was supported by the Deutsche Forschungsgemeinschaft (DFG, German Research Foundation) under grant 396021762 -- TRR 257 ``Particle Physics Phenomenology after the Higgs Discovery". This publication is part of the project ``Beauty decays: the quest for extreme precision'' of the Open Competition Domain Science which is financed by the Dutch Research Council (NWO).

\appendix
\section{Input parameters} \label{app:inputs}
For the HQE elements, we use the Reparametrisation Invariant (RPI) basis for the $q^2$ moments and the historical basis (also called ``perp'' basis in some literature) for the $E_\ell$ moments (see \cite{Mannel:2010wj, Gambino:2016jkc} for their definitions). Here we list the definitions for the HQE parameters in the RPI basis up to $1/m_b^3$
\begin{align}
    2m_B\mu_\pi^2&=-\langle \bar{b}_v\, (iD)^2\, b_v\rangle\ ,\nonumber \\
    2m_B\mu_G^2&=\langle \bar{b}_v\, (iD_\alpha)\,(iD_\beta)\,(-i\sigma^{\alpha\beta})\,b_v\rangle\ ,\nonumber\\
    2m_B\tilde{\rho}_D^3&=\frac{1}{2}\langle \bar{b}_v\,\left[(iD_\mu),\Big[\Big((ivD)+\frac{1}{2m_b}(iD)^2\Big),(iD^\mu)\Big]\right]\,b_v\rangle\ ,\nonumber\\
    2m_B\rho_{LS}^3&=\frac{1}{2}\langle \bar{b}_v\,\left[(iD_\alpha),\Big[(ivD),(iD_\beta)\Big]\right](-i\sigma^{\alpha\beta})\,b_v\rangle  \ .
\end{align}
For the definitions of the HQE parameters at $1/m_b^4$ and $1/m_b^5$ in the RPI basis, we refer to \cite{Mannel:2023yqf, Mannel:2018mqv}.

In Table \ref{tab:inputs}, we present the input values for the phenomenological predictions presented in this paper. To obtain estimates for the HQE parameters, we use the ``lowest-lying state saturation ansatz'' (LLSA) \cite{Heinonen:2014dxa}. This ansatz allows us to express the HQE elements in terms of the excitation energies $\epsilon_{1/2},\  \epsilon_{3/2}$ and the $1/m_b^2$ elements $(\mu_\pi^2)^\perp$ and $(\mu_G^2)^\perp$. For the LLSA expressions for HQE elements up to $1/m_b^5$ were recently discussed in \cite{Mannel:2023yqf}. Using those and the inputs in Table~\ref{tab:inputs}, we find the estimates in Table~\ref{tab:LLSA} for the HQE parameters in both the historical basis and the RPI basis. The conversion between these two bases is discussed in \cite{Mannel:2023yqf}.

Since $\hqe^n$ is supposed to set the scale of the HQE parameters at dimension $n+3$, as introduced in \eqref{eq:traceform}, we might take the $n^{\rm{th}}$ root of the LLSA approximations for the HQE parameters at dimension $n+3$, which is expected to be $\sim\hqe$. If we average the roots found from Table \ref{tab:LLSA}, we find a value of $\sim0.5$ GeV. Therefore, we use that as a default value for $\hqe$ in throughout this paper. 
\begin{table}[t!]
    \centering
    \begin{tabular}{|l||ll|}
    \multicolumn{3}{c}{\textbf{Input values}}\\
    \hline
        $m_b^{kin}$ &4.573  {\rm GeV} &\cite{Bordone:2021oof}\\
        $\overline{m}_c(2\ \text{GeV})$ & 1.092  {\rm GeV}&\cite{Bordone:2021oof}\\
        $m_B$  & 5.279 GeV & \cite{Workman:2022ynf}\\
    $\epsilon_{1/2}$&0.390  {\rm GeV}&\cite{Heinonen:2014dxa}\\
        $\epsilon_{3/2}$&0.476  {\rm GeV}&\cite{Heinonen:2014dxa}\\
        $(\mu_\pi^2)^\perp$&0.477  {\rm GeV}$^2$&\cite{Bordone:2021oof}\\
        $(\mu_G^2)^\perp$ &0.306  {\rm GeV}$^2$&\cite{Bordone:2021oof}\\
        \hline
    \end{tabular} 
    \caption{The input values used for our numerical analysis.}
    \label{tab:inputs}
\end{table} 

\begin{table}[t!]
    \centering
    \begin{tabular}{|l||l|}
     \multicolumn{2}{c}{\textbf{LLSA approximation}}\\
     \multicolumn{2}{c}{\textbf{Historical basis}}\\
        \hline
        $(\rho_D^3)^\perp$ & 0.232 {\rm GeV}$^3$\\
        $(\rho_{LS}^3)^\perp$ & -0.161 {\rm GeV}$^3$\\
        \hline
        $m_1$ & 0.126 {\rm GeV}$^4$\\
        $m_2$ & -0.112 {\rm GeV}$^4$\\
        $m_3$ & -0.062 {\rm GeV}$^4$\\
        $m_4$ & 0.397 {\rm GeV}$^4$\\
        $m_5$ & 0.081 {\rm GeV}$^4$\\
        $m_6$ & 0.062 {\rm GeV}$^4$\\
        $m_7$ & -0.039 {\rm GeV}$^4$\\
        $m_8$ & -1.17 {\rm GeV}$^4$\\
        $m_9$ & -0.393 {\rm GeV}$^4$\\
        \hline
    \end{tabular} \hspace{0.5cm}
    \begin{tabular}{|l||l|}
    \multicolumn{2}{c}{\textbf{LLSA approximation}}\\
    \multicolumn{2}{c}{\textbf{Historical basis}}\\
        \hline 
        $r_1$& 0.049 {\rm GeV}$^5$\\
        $r_2$& -0.106 {\rm GeV}$^5$\\
        $r_3$& -0.027 {\rm GeV}$^5$\\
        $r_4$& -0.043 {\rm GeV}$^5$\\
        $r_5$& 0.00 {\rm GeV}$^5$\\
        $r_6$& 0.00 {\rm GeV}$^5$\\
        $r_7$& 0.00 {\rm GeV}$^5$\\
        $r_8$& -0.039 {\rm GeV}$^5$\\
        $r_9$& 0.074 {\rm GeV}$^5$\\
        $r_{10}$& 0.068 {\rm GeV}$^5$\\
        $r_{11}$& 0.0059 {\rm GeV}$^5$\\
        $r_{12}$& 0.010 {\rm GeV}$^5$\\
        $r_{13}$& -0.055 {\rm GeV}$^5$\\
        $r_{14}$& 0.039 {\rm GeV}$^5$\\
        $r_{15}$& 0.00 {\rm GeV}$^5$\\
        $r_{16}$& 0.00 {\rm GeV}$^5$\\
        $r_{17}$& 0.00 {\rm GeV}$^5$\\
        $r_{18}$& 0.00 {\rm GeV}$^5$\\
        \hline
    \end{tabular}\hspace{0.5cm}
    \begin{tabular}{|l||l|}
     \multicolumn{2}{c}{\textbf{LLSA approximation}}\\
     \multicolumn{2}{c}{\textbf{RPI-basis}}\\
        \hline
        $\mu_\pi^2$&0.477 {\rm GeV}$^2$\\
        $\mu_G^2$&0.290 {\rm GeV}$^2$\\
        \hline
        $\tilde{\rho}_D^3$ &0.205 {\rm GeV}$^3$\\
        \hline
        $\tilde{r}_E^4$ &0.098 {\rm GeV}$^4$\\
        $r_G^4$ &0.16 {\rm GeV}$^4$\\
        $\tilde{s}_E^4$ &-0.074 {\rm GeV}$^4$\\
        $s_B^4$&-0.14 {\rm GeV}$^4$\\
        $s_{qB}^4$ &-1.00 {\rm GeV}$^4$\\
        \hline 
        $X_1^5$&0.049 {\rm GeV}$^5$\\
        $X_2^5$ &0.00  {\rm GeV}$^5$\\
        $X_3^5$ &0.094  {\rm GeV}$^5$\\
        $X_4^5$ &-0.41 {\rm GeV}$^5$\\
        $X_5^5$ &-0.039 {\rm GeV}$^5$\\
        $X_6^5$ &0.00  {\rm GeV}$^5$\\
        $X_7^5$&0.091 {\rm GeV}$^5$\\
        $X_8^5$&-0.0030 {\rm GeV}$^5$\\
        $X_9^5$&0.27 {\rm GeV}$^5$\\
        $X_{10}^5$ &0.025 {\rm GeV}$^5$\\
        \hline
    \end{tabular} 
    \caption{Estimates for the HQE parameters in the historical basis and the RPI basis based on the LLSA approximation using the input values from Table~\ref{tab:inputs}.}
    \label{tab:LLSA}
\end{table}

\bibliographystyle{jhep} 
\bibliography{refs.bib} 

\providecommand{\href}[2]{#2}\begingroup\raggedright\begin{thebibliography}{10}

\bibitem{Poggio:1975af}
E.~C. Poggio, H.~R. Quinn and S.~Weinberg, \emph{{Smearing the Quark Model}},
  \href{http://dx.doi.org/10.1103/PhysRevD.13.1958}{\emph{Phys. Rev. D} {\bf
  13} (1976) 1958}.

\bibitem{Shifman:2003de}
M.~Shifman, \emph{{The quark hadron duality}}, {\emph{eConf} {\bf C030614}
  (2003) 001}.

\bibitem{Bigi:2002fj}
I.~I. Bigi and T.~Mannel, \emph{{Parton hadron duality in B meson decays}},  in
  \emph{{Workshop on CKM Unitarity Triangle (CERN 2002-2003)}}, pp.~61--67, 12,
  2002.
\newblock \href{http://arxiv.org/abs/hep-ph/0212021}{{\tt hep-ph/0212021}}.

\bibitem{Bigi:2001ys}
I.~I.~Y. Bigi and N.~Uraltsev, \emph{{A Vademecum on quark hadron duality}},
  \href{http://dx.doi.org/10.1142/S0217751X01005535}{\emph{Int. J. Mod. Phys.
  A} {\bf 16} (2001) 5201--5248},
  [\href{http://arxiv.org/abs/hep-ph/0106346}{{\tt hep-ph/0106346}}].

\bibitem{Shifman:2000jv}
M.~A. Shifman, \emph{{Quark hadron duality}},  in \emph{{8th International
  Symposium on Heavy Flavor Physics}}, vol.~3, (Singapore), pp.~1447--1494,
  World Scientific, 7, 2000.
\newblock \href{http://arxiv.org/abs/hep-ph/0009131}{{\tt hep-ph/0009131}}.
\newblock \href{http://dx.doi.org/10.1142/9789812810458_0032}{DOI}.

\bibitem{Blok:1997hs}
B.~Blok, M.~A. Shifman and D.-X. Zhang, \emph{{An Illustrative example of how
  quark hadron duality might work}},
  \href{http://dx.doi.org/10.1103/PhysRevD.57.2691}{\emph{Phys. Rev. D} {\bf
  57} (1998) 2691--2700}, [\href{http://arxiv.org/abs/hep-ph/9709333}{{\tt
  hep-ph/9709333}}].

\bibitem{Chibisov:1996wf}
B.~Chibisov, R.~D. Dikeman, M.~A. Shifman and N.~Uraltsev, \emph{{Operator
  product expansion, heavy quarks, QCD duality and its violations}},
  \href{http://dx.doi.org/10.1142/S0217751X97001316}{\emph{Int. J. Mod. Phys.
  A} {\bf 12} (1997) 2075--2133},
  [\href{http://arxiv.org/abs/hep-ph/9605465}{{\tt hep-ph/9605465}}].

\bibitem{Benson:2003kp}
D.~Benson, I.~I. Bigi, T.~Mannel and N.~Uraltsev, \emph{{Imprecated, yet
  impeccable: On the theoretical evaluation of Gamma(B ---\ensuremath{>} X(c) l
  nu)}}, \href{http://dx.doi.org/10.1016/S0550-3213(03)00452-8}{\emph{Nucl.
  Phys. B} {\bf 665} (2003) 367--401},
  [\href{http://arxiv.org/abs/hep-ph/0302262}{{\tt hep-ph/0302262}}].

\bibitem{Bigi:1996si}
I.~I.~Y. Bigi, M.~A. Shifman, N.~Uraltsev and A.~I. Vainshtein, \emph{{High
  power $n$ of $m_b$ in beauty widths and $n=5\to \infty $ limit}},
  \href{http://dx.doi.org/10.1103/PhysRevD.56.4017}{\emph{Phys. Rev. D} {\bf
  56} (1997) 4017--4030}, [\href{http://arxiv.org/abs/hep-ph/9704245}{{\tt
  hep-ph/9704245}}].

\bibitem{Dassinger:2006md}
B.~M. Dassinger, T.~Mannel and S.~Turczyk, \emph{{Inclusive semi-leptonic B
  decays to order $1/m_b^4$}},
  \href{http://dx.doi.org/10.1088/1126-6708/2007/03/087}{\emph{JHEP} {\bf 03}
  (2007) 087}, [\href{http://arxiv.org/abs/hep-ph/0611168}{{\tt
  hep-ph/0611168}}].

\bibitem{Mannel:2010wj}
T.~Mannel, S.~Turczyk and N.~Uraltsev, \emph{{Higher Order Power Corrections in
  Inclusive B Decays}},
  \href{http://dx.doi.org/10.1007/JHEP11(2010)109}{\emph{JHEP} {\bf 11} (2010)
  109}, [\href{http://arxiv.org/abs/1009.4622}{{\tt 1009.4622}}].

\bibitem{Mannel:2023yqf}
T.~Mannel, I.~S. Milutin and K.~K. Vos, \emph{{Inclusive semileptonic $ b\to
  c\ell \overline{\nu} $ decays to order $ 1/{m}_b^5 $}},
  \href{http://dx.doi.org/10.1007/JHEP02(2024)226}{\emph{JHEP} {\bf 02} (2024)
  226}, [\href{http://arxiv.org/abs/2311.12002}{{\tt 2311.12002}}].

\bibitem{Fael:2020tow}
M.~Fael, K.~Sch\"onwald and M.~Steinhauser, \emph{{Third order corrections to
  the semileptonic b\textrightarrow{}c and the muon decays}},
  \href{http://dx.doi.org/10.1103/PhysRevD.104.016003}{\emph{Phys. Rev. D} {\bf
  104} (2021) 016003}, [\href{http://arxiv.org/abs/2011.13654}{{\tt
  2011.13654}}].

\bibitem{Fael:2022frj}
M.~Fael, K.~Sch\"onwald and M.~Steinhauser, \emph{{A first glance to the
  kinematic moments of B \textrightarrow{}
  X$_{c}$\ensuremath{\ell}\ensuremath{\nu} at third order}},
  \href{http://dx.doi.org/10.1007/JHEP08(2022)039}{\emph{JHEP} {\bf 08} (2022)
  039}, [\href{http://arxiv.org/abs/2205.03410}{{\tt 2205.03410}}].

\bibitem{Becher:2007tk}
T.~Becher, H.~Boos and E.~Lunghi, \emph{{Kinetic corrections to $B \to X_{c}
  \ell \bar{\nu}$ at one loop}},
  \href{http://dx.doi.org/10.1088/1126-6708/2007/12/062}{\emph{JHEP} {\bf 12}
  (2007) 062}, [\href{http://arxiv.org/abs/0708.0855}{{\tt 0708.0855}}].

\bibitem{Alberti:2012dn}
A.~Alberti, T.~Ewerth, P.~Gambino and S.~Nandi, \emph{{Kinetic operator effects
  in $\bar{B}\to X_c l \nu$ at O($\alpha_s$)}},
  \href{http://dx.doi.org/10.1016/j.nuclphysb.2013.01.005}{\emph{Nucl. Phys. B}
  {\bf 870} (2013) 16--29}, [\href{http://arxiv.org/abs/1212.5082}{{\tt
  1212.5082}}].

\bibitem{Alberti:2013kxa}
A.~Alberti, P.~Gambino and S.~Nandi, \emph{{Perturbative corrections to power
  suppressed effects in semileptonic B decays}},
  \href{http://dx.doi.org/10.1007/JHEP01(2014)147}{\emph{JHEP} {\bf 01} (2014)
  147}, [\href{http://arxiv.org/abs/1311.7381}{{\tt 1311.7381}}].

\bibitem{Mannel:2021zzr}
T.~Mannel, D.~Moreno and A.~A. Pivovarov, \emph{{NLO QCD corrections to
  inclusive $b \rightarrow c \ell \bar{\nu}$decay spectra up to~$1/m_Q^3$}},
  \href{http://dx.doi.org/10.1103/PhysRevD.105.054033}{\emph{Phys. Rev. D} {\bf
  105} (2022) 054033}, [\href{http://arxiv.org/abs/2112.03875}{{\tt
  2112.03875}}].

\bibitem{Gambino:2016jkc}
P.~Gambino, K.~J. Healey and S.~Turczyk, \emph{{Taming the higher power
  corrections in semileptonic B decays}},
  \href{http://dx.doi.org/10.1016/j.physletb.2016.10.023}{\emph{Phys. Lett. B}
  {\bf 763} (2016) 60--65}, [\href{http://arxiv.org/abs/1606.06174}{{\tt
  1606.06174}}].

\bibitem{Bordone:2021oof}
M.~Bordone, B.~Capdevila and P.~Gambino, \emph{{Three loop calculations and
  inclusive Vcb}},
  \href{http://dx.doi.org/10.1016/j.physletb.2021.136679}{\emph{Phys. Lett. B}
  {\bf 822} (2021) 136679}, [\href{http://arxiv.org/abs/2107.00604}{{\tt
  2107.00604}}].

\bibitem{Bernlochner:2022ucr}
F.~Bernlochner, M.~Fael, K.~Olschewsky, E.~Persson, R.~van Tonder, K.~K. Vos
  et~al., \emph{{First extraction of inclusive V$_{cb}$ from q$^{2}$ moments}},
  \href{http://dx.doi.org/10.1007/JHEP10(2022)068}{\emph{JHEP} {\bf 10} (2022)
  068}, [\href{http://arxiv.org/abs/2205.10274}{{\tt 2205.10274}}].

\bibitem{Finauri:2023kte}
G.~Finauri and P.~Gambino, \emph{{The q$^{2}$ moments in inclusive semileptonic
  B decays}}, \href{http://dx.doi.org/10.1007/JHEP02(2024)206}{\emph{JHEP} {\bf
  02} (2024) 206}, [\href{http://arxiv.org/abs/2310.20324}{{\tt 2310.20324}}].

\bibitem{Heinonen:2014dxa}
J.~Heinonen and T.~Mannel, \emph{{Improved Estimates for the Parameters of the
  Heavy Quark Expansion}},
  \href{http://dx.doi.org/10.1016/j.nuclphysb.2014.09.017}{\emph{Nucl. Phys. B}
  {\bf 889} (2014) 46--63}, [\href{http://arxiv.org/abs/1407.4384}{{\tt
  1407.4384}}].

\bibitem{Chay:1994si}
J.~Chay and S.-J. Rey, \emph{{Instanton contribution to B ---\ensuremath{>}
  X(mu) e anti-neutrino decay}},
  \href{http://dx.doi.org/10.1007/BF01620717}{\emph{Z. Phys. C} {\bf 68} (1995)
  431--438}, [\href{http://arxiv.org/abs/hep-ph/9404214}{{\tt
  hep-ph/9404214}}].

\bibitem{Falk:1995yc}
A.~F. Falk and A.~Kyatkin, \emph{{Instantons and the endpoint of the lepton
  energy spectrum in charmless semileptonic B decays}},
  \href{http://dx.doi.org/10.1103/PhysRevD.52.5049}{\emph{Phys. Rev. D} {\bf
  52} (1995) 5049--5055}, [\href{http://arxiv.org/abs/hep-ph/9502248}{{\tt
  hep-ph/9502248}}].

\bibitem{Herren:2022spb}
F.~Herren, \emph{{The forward-backward asymmetry and differences of partial
  moments in inclusive semileptonic $B$ decays}},
  \href{http://dx.doi.org/10.21468/SciPostPhys.14.2.020}{\emph{SciPost Phys.}
  {\bf 14} (2023) 020}, [\href{http://arxiv.org/abs/2205.03427}{{\tt
  2205.03427}}].

\bibitem{Belle:2021idw}
{\scshape Belle} collaboration, R.~van Tonder et~al., \emph{{Measurements of
  $q^2$ Moments of Inclusive $B \rightarrow X_c \ell^+ \nu_{\ell}$ Decays with
  Hadronic Tagging}},
  \href{http://dx.doi.org/10.1103/PhysRevD.104.112011}{\emph{Phys. Rev. D} {\bf
  104} (2021) 112011}, [\href{http://arxiv.org/abs/2109.01685}{{\tt
  2109.01685}}].

\bibitem{Belle-II:2022evt}
{\scshape Belle-II} collaboration, F.~Abudin\'en et~al., \emph{{Measurement of
  lepton mass squared moments in
  B\textrightarrow{}Xc\ensuremath{\ell}\ensuremath{\nu}\textasciimacron{}\ensuremath{\ell}
  decays with the Belle II experiment}},
  \href{http://dx.doi.org/10.1103/PhysRevD.107.072002}{\emph{Phys. Rev. D} {\bf
  107} (2023) 072002}, [\href{http://arxiv.org/abs/2205.06372}{{\tt
  2205.06372}}].

\bibitem{Mannel:2018mqv}
T.~Mannel and K.~K. Vos, \emph{{Reparametrization Invariance and Partial
  Re-Summations of the Heavy Quark Expansion}},
  \href{http://dx.doi.org/10.1007/JHEP06(2018)115}{\emph{JHEP} {\bf 06} (2018)
  115}, [\href{http://arxiv.org/abs/1802.09409}{{\tt 1802.09409}}].

\bibitem{Workman:2022ynf}
{\scshape Particle Data Group} collaboration, R.~L. Workman and Others,
  \emph{{Review of Particle Physics}},
  \href{http://dx.doi.org/10.1093/ptep/ptac097}{\emph{PTEP} {\bf 2022} (2022)
  083C01}.

\end{thebibliography}\endgroup

\end{document}